\documentclass[useAMS, usenatbib]{mn2e}
\input psfig.sty

\newcommand {\apj} {ApJ}
\newcommand {\apjl} {ApJL}
\newcommand {\apjs} {ApJS}
\newcommand {\mnras} {MNRAS}

\newcommand {\aap} {A\&A}
\newcommand {\aj} {AJ}

\newcommand {\araa} {ARA\&A}

\newcommand {\etal} {et~al.~}
\def \spose#1{\hbox  to 0pt{#1\hss}}  
\newcommand {\lta} {\mathrel{\spose{\lower 3pt\hbox{$\sim$}}\raise  2.0pt\hbox{$<$}}}
\newcommand {\gta} {\mathrel{\spose{\lower  3pt\hbox{$\sim$}}\raise 2.0pt\hbox{$>$}}}

\newcommand {\kms} {\ifmmode  \,\rm km\,s^{-1} \else $\,\rm km\,s^{-1}  $ \fi }
\newcommand {\kpc} {\ifmmode  {\rm kpc}  \else ${\rm  kpc}$ \fi  }  
\newcommand {\Msun} {\ifmmode M_{\odot} \else $M_{\odot}$ \fi} 
\newcommand {\hMsun} {\ifmmode h^{-1}\,\rm M_{\odot} \else $h^{-1}\,\rm M_{\odot}$ \fi}

\newcommand {\LCDM} {\ifmmode \Lambda{\rm CDM} \else $\Lambda{\rm CDM}$ \fi}

\newcommand {\Rvir} {\ifmmode R_{\rm vir} \else $R_{\rm vir}$ \fi}
\newcommand {\Vvir} {\ifmmode V_{\rm  vir} \else  $V_{\rm vir}$  \fi} 
\newcommand {\Mvir} {\ifmmode M_{\rm  vir} \else $M_{\rm  vir}$ \fi}  
\newcommand {\Jvir} {\ifmmode J_{\rm vir} \else $J_{\rm vir}$ \fi} 
\newcommand {\Evir} {\ifmmode E_{\rm vir} \else $E_{\rm vir}$ \fi} 
\newcommand {\lamgal} {\ifmmode \lambda_{\rm gal}  \else $\lambda_{\rm gal}$ \fi} 
\newcommand {\lamstar} {\ifmmode \lambda_{\rm star}  \else $\lambda_{\rm star}$ \fi} 
\newcommand {\lam} {\ifmmode \lambda  \else $\lambda$ \fi} 

\newcommand {\mgal}    {\ifmmode m_{\rm gal}    \else $m_{\rm gal}$ \fi} 
\newcommand {\Mstar}    {\ifmmode M_{\rm star}    \else $M_{\rm star}$ \fi} 
\newcommand {\Mgal}  {\ifmmode M_{\rm gal}  \else $M_{\rm gal}$ \fi}
\newcommand {\jgal} {\ifmmode j_{\rm gal} \else $j_{\rm gal}$ \fi}  
\newcommand {\Jgal} {\ifmmode J_{\rm gal} \else $J_{\rm gal}$ \fi}  
\newcommand {\Vrot} {\ifmmode  V_{\rm rot} \else $V_{\rm rot}$ \fi} 

\title[On the Origin of Exponential Galaxy Disks]
      {On the Origin of Exponential Galaxy Disks}
    
\author[Dutton]
       {Aaron A. Dutton$^{1}$\thanks{dutton@ucolick.org}\\
      $^1$UCO/Lick Observatory, University of California, Santa Cruz, CA 95064\\}

\begin{document}
             
\date{submitted to MNRAS}
             
\pagerange{\pageref{firstpage}--\pageref{lastpage}}\pubyear{2008}

\maketitle           

\label{firstpage}


\begin{abstract}
  One  of the most  important unresolved  issues for  galaxy formation
  theory is  to understand the  origin of exponential  galaxy disks.
  We use a disk galaxy evolution model to investigate whether galaxies
  with exponential  surface brightness profiles  can be produced  in a
  cosmologically motivated  framework for disk  galaxy formation.  Our
  model follows the accretion, cooling, and ejection of baryonic mass,
  as a  function of  radius, inside growing  dark matter  haloes.  The
  surface  density  profile of  the  disk  is  determined by  detailed
  angular  momentum conservation,  starting from  the  distribution of
  specific angular momentum as found in cosmological simulations.
  Exponential and quasi-exponential disks can be produced by our model
  through a  combination of supernova driven  galactic outflows (which
  preferentially  remove  low  angular momentum  material),  intrinsic
  variation in the angular momentum  distribution of the halo gas, and
  the  inefficiency  of  star   formation  at  large  radii.  
  We use observations  from the SDSS NYU-VAGC to  show that the median
  S\'ersic index of late-type galaxies is a strong function of stellar
  mass.   For blue  galaxies, low  mass galaxies  have  $n\simeq 1.3$,
  while high mass galaxies have $n\simeq 4$, with a transition mass of
  $\Mstar  \simeq  2.5\times  10^{10}\Msun$.   Our model  with  energy
  driven outflows correctly reproduces  this trend, whereas our models
  with  momentum driven  outflows  and no  outflows  over predict  the
  S\'ersic  indices in low  mass galaxies. 
  We  show that  the observed  fraction of  ``bulge-less'' exponential
  galaxies is a  strong function of stellar mass.   For Milky-Way mass
  galaxies ($\Vrot \simeq 220  \kms, \Mstar \simeq 10^{11}\Msun$) less
  than 0.1\%  of blue  galaxies are bulge-less,  whereas for  M33 mass
  galaxies  ($  \Vrot \simeq  120  \kms,  \Mstar \simeq  10^{10}\Msun$
  bulge-less and quasi-bulge-less galaxies  are more common, with 45\%
  of  blue  galaxies having  S\'ersic  index  $n<1.5$.  These  results
  suggest  that the  difficulty  of hierarchical  formation models  to
  produce  bulge-less  Milky-Way mass  galaxies  is,  in  fact, not  a
  problem.   However,  the  problem  of producing  M33  like  galaxies
  remains,  and  will  provide  a  key test  for  hierarchical  galaxy
  formation models, and feedback models inparticular.

\end{abstract}


\setcounter{footnote}{1}


\section{Introduction}
\label{sec:intro}
In the  current paradigm for galaxy  formation, set forth  by White \&
Rees  (1978)  and  Fall   \&  Efstathiou  (1980),  disk  galaxies  are
considered to form  through the cooling of baryons  inside dark matter
haloes  that grow by  means of  gravitational instability  and acquire
angular  momentum from  cosmological  torques.  In  this paradigm  the
baryons acquire the same  distribution of specific angular momentum as
the  dark matter,  and  this is  conserved  during cooling.   Analytic
models based on these assumptions have been able to produce disk sizes
that are  in reasonable agreement with  observations (e.g.  Dalcanton,
Spergel \& Summers 1997; Mo, Mao \& White 1998; de Jong \& Lacey 2000;
Dutton \etal 2007).   In this picture the large  scatter in disk sizes
(including  the existence  of low  surface brightness  galaxies)  is a
natural consequence of the large scatter in halo spin parameter.

However, detailed  hydro-dynamical simulations aimed  at investigating
the process of galaxy formation have indicated a potentially important
problem  for the  cold dark  matter (CDM)  model. In  CDM cosmologies,
haloes  form  hierarchically  by   the  merging  of  many  lower  mass
haloes.  Because  the  cooling  in  dense,  low-mass  haloes  is  very
efficient, the  baryons in  these systems have  already cooled  by the
time they  merge with  the more massive  protogalaxy.  They  reach the
center of the  potential well by means of  dynamical friction, through
which they  lose a significant  fraction of their angular  momentum to
the  dark   matter.  Consequently,  the  disks  that   form  in  these
simulations are an order of magnitude too small (Navarro \& Benz 1991;
Navarro \&  White 1994; Navarro \&  Steinmetz 2000). This  is known as
angular momentum catastrophe.

Attempted solutions to this problem include stellar feedback and/or
ionizing background radiation which can prevent the cooling of the gas
in low-mass haloes, thereby reducing the angular momentum loss (e.g.
Dom\'inguez-Tenreiro, Tissera \& S\'aiz 1998; Weil, Eke \& Efstathiou
1998; Sommer-Larsen, Gelato \& Vedel 1999; Eke, Efstathiou \& Wright
2000; Maller \& Dekel 2002). An alternative suggestion has been to
reduce the number of low mass haloes by reducing the amplitude of the
power spectrum on small scales (e.g. Kamionkowski \& Liddle 2000;
Sommer-Larsen \& Dolgov 2001).  However, these effects have only
partially successful in resolving the angular momentum catastrophe.
More recent numerical simulations have shown that in addition to
feedback, very high numerical resolution is also required in order to
avoid numerical angular momentum losses (e.g. Kaufmann \etal 2007;
Ceverino \& Klypin 2007; Mayer \etal 2008).

A second problem, related to angular momentum, is to understand the
exponential density profile of stellar galactic disks.  If disk
galaxies form as envisaged in the standard picture their resulting
surface density distribution is directly related to the specific
angular momentum distribution of the protogalaxy.  Motivated by the
work of Mestel (1963), Crampin \& Hoyle (1964) and Gunn (1982),
Dalcanton \etal (1997) made the assumption that the protogalaxy has
the AMD of a uniform sphere in solid body rotation, and showed that
the resulting disks are more concentrated than exponential.

Firmani \& Avila-Reese (2000) and van den Bosch (2001) refined this
model by following the build-up and evolution of the disk density
profile within a (cosmologically motivated) growing dark matter
halo. In these models each shell of collapsing material is assumed to
have the angular momentum distribution of a shell in solid body
rotation, and the normalization of the angular momentum distribution
is calculated by assuming that the spin parameter of the halo is
independent of redshift. In the Firmani \& Avila-Reese (2000) model
the baryonic mass fraction of the galaxy was treated as a free
parameter, with a fiducial value of 0.05.  In the standard CDM
cosmology (i.e. $\Omega_{\rm m,0}=1$), such a baryon fraction is
consistent with the universal value. However, with the currently
favored a \LCDM cosmology (where $\Omega_{\rm m,0} \simeq 0.26$) the
baryon fraction is $\simeq 0.17$, and the low baryons fractions in
observed galaxies require an explanation. In the van den Bosch (2001)
model the galaxy mass fraction is determined by the efficiencies of
cooling and supernova (SN) driven outflows. Cooling is very efficient in
galaxy mass halos, thus in order to produce the low galaxy mass
fractions that are observed (e.g. Hoekstra \etal 2005; Mandelbaum
\etal 2006) and required to reconcile the stellar and halo mass
functions (e.g. Yang \etal 2007; Conroy \& Wechsler 2008) galactic
outflows are needed.

The van den Bosch (2001) model highlights the problem of producing
pure exponential stellar galactic disks in a cosmological framework.
In these models the disk may be approximately exponential over a few
scale lengths, but the disk is invariably more concentrated than
exponential in the center. A potential short-coming of the Firmani \&
Avila-Reese (2000) and van den Bosch (2001) models is the assumption
that the accreting mass shells have the AMDs of shells in solid body
rotation.

Bullock \etal (2001b) determined the distribution of specific angular
momentum in \LCDM haloes from cosmological N-body simulations, and again
concluded that these distributions will form overly concentrated
disks. This excess of low angular momentum material has been confirmed
by a number of subsequent studies (van den Bosch \etal 2002; Chen
\etal 2003; Sharma \& Steinmetz 2005). These simulations also
indicated that CDM haloes have an excess of high specific angular
momentum material, relative to that required to make an exponential
disk. Furthermore, not even a single bulge-less disk galaxy with an
exponential stellar density profile has been formed in hydrodynamical
simulations in the \LCDM paradigm. Thus one of the most important
unresolved issues in galaxy formation is to understand the origin of
the exponential profile of stellar galactic disks.

There are several reasons why the specific angular momentum profile of
the stellar disk may be different than that of the dark matter and hot
gas. Firstly only a fraction of the hot gas will be able to cool;
secondly some of the cooled gas may be ejected from the galaxy or
progenitor galaxies (e.g. Maller \& Dekel 2002) via SN driven
winds; thirdly only a fraction of the cooled gas will be converted
into stars. Furthermore the surface brightness profile in optical
light may not reliably trace the surface density profile in stellar
mass.  Thus an important question is whether the processes of star
formation, feedback and stellar populations can produce exponential
surface brightness profiles.  If this is not possible then it signals
that the cold baryons acquire a different distribution of specific
angular momentum than the dark matter, or more fundamentally signals a
failure of the CDM paradigm of structure formation.

In this paper we use the models described in Dutton \& van den Bosch
(2008), which are an updated version of the models used by van den
Bosch (2001, 2002) to investigate the origin of the exponential light
profiles of galactic disks. This model includes the detailed cooling,
star formation and ejection of gas, as a function of radius, inside
growing dark matter haloes with cosmologically motivated density
profiles and angular momentum distributions.

This paper is organized as follows.  In \S 2 we give an overview of
the models, in \S 3 we discuss the impact of the initial angular
momentum distribution, star formation, feedback, halo mass, adiabatic
contraction and stellar populations on the structure of galaxy
disks. In \S 4 we use observations from the SDSS to determine how
frequent bulge-less galaxy disks are, and in \S 5 we make a comparison
between these observations and Monte Carlo model samples.  We discuss
our results in \S 6 and give a summary in \S 7.


\section{Disk Galaxy Evolution Models}
\label{sec:models}
Here we give a brief overview  of the disk galaxy evolution model used
in this paper. This model is  described in detail in Dutton \& van den
Bosch (2008;  hereafter DB08).  The  key difference with  almost all
disk evolution  models is that  in this model  the inflow (due  to gas
cooling), outflow (due to SN  driven winds), star formation rates, and
metallicity, are computed  {\it as a function of  radius}, rather than
being  treated  as  global  parameters.   The  main  assumptions  that
characterize the framework of our models are the following:
\begin{enumerate}
\item  Dark matter  haloes around  disk  galaxies grow  by the  smooth
  accretion of mass which we model with the Wechsler \etal (2002) mass
  accretion history  (MAH). The shape of  this MAH is  specified by the
  concentration of the halo at redshift zero;
\item The structure of the halo is given by the NFW profile (Navarro,
  Frenk, \& White 1997), which is specified by two parameters: the
  mass and concentration. The evolution of the concentration parameter
  is given by the Bullock \etal (2001a) model with parameters
  (Macci\`o \etal 2008) for a WMAP 5th year cosmology (Dunkley \etal
  2009);
\item  Gas  that  enters  the  halo  is shock  heated  to  the  virial
  temperature, and acquires the  same distribution of specific angular
  momentum  as   the  dark  matter.   We  use  the   angular  momentum
  distributions of the halo  as parametrized by Bullock \etal (2001b)
  and Sharma \& Steinmetz (2005);
\item Gas cools radiatively, conserving its specific angular momentum,
  and forms a disk in centrifugal equilibrium;
\item Star formation occurs according  to a Schmidt (1959) type law on
  the  dense  molecular gas,  which  is  computed  following Blitz  \&
  Rosolowsky (2006);
\item SN  feedback re-heats some  of the cold gas,  ejecting it
  from the disk and halo;
\item Stars eject metals into  the inter stellar medium, enriching the
  cold gas.
\item Bruzual  \& Charlot  (2003) stellar population  synthesis models
  are convolved with the star formation histories and metallicities to
  derive surface brightness.
\end{enumerate}
Each model galaxy  consists of five mass components:  dark matter, hot
halo gas, disk mass in stars,  disk mass in cold gas, and ejected gas.
The  dark matter  and the  hot gas  are assumed  to be  distributed in
spherical shells,  the disk mass  is assumed to be  in infinitesimally
thin annuli.  Throughout this paper we  refer to $R$ as radius, $t$ as
time (where $t=0$ is defined as the Big Bang) and $z$ as redshift.

For each galaxy we set up a radial grid with 200 bins quadratically
sampled from between 0.001 and 1 times the redshift zero virial
radius.  As described in DB08 for the purpose of conserving angular
momentum it is more convenient to use a grid in specific angular
momentum, $j$.  We convert the initial grid in $R$, to a grid in $j$
using $j^2/G = R M(R)$, where $M(R)$ is the halo mass enclosed within
a sphere of radius $R$.  We follow the formation and evolution of the
galaxy using 400 redshift steps, quadratically sampled from $z=10$ to
$z=0$.  For each time step we compute the changes in the various mass
components in each radial bin.

\subsection{Angular Momentum Distribution}
The radius  at which the  cooled gas ends  up depends on  its specific
angular  momentum,  $j$.   Van  den  Bosch (2001;  2002)  assumed  the
$j$-distribution to  be that  of a shell  in solid body  rotation.  In
this paper we adopt cosmologically motivated specific angular momentum
distributions (AMD).

The AMD can be specified  by 2 parameters: a normalization ($\lambda$)
and a  shape ($\mu$,  or $\alpha$). Both  the normalization  and shape
parameters are log-normally distributed, with significant scatter.  We
assume that  the spin and shape parameters  are uncorrelated, although
Bullock  \etal (2001b)  show  that  there may  be  a weak  correlation
between $\mu$ and $\lambda$, which needs to be investigated further.

The spin parameter is defined by
\begin{equation}
\lambda = \frac{\Jvir | \Evir |^{1/2}}{G\Mvir^{5/2}},
\end{equation}
where $\Mvir, \Jvir$, and $\Evir$ are the mass, total angular momentum
and energy of the halo, respectively.

Bullock \etal (2001b; hereafter B01) showed that dark matter haloes in
a  \LCDM cosmology have  specific angular  momentum profiles  that are
well fitted by
\begin{equation}
\label{eqn:mjbul}
m(j)=\frac{M(<j)}{\Mvir} = \mu \frac{j/j_{\rm max}} {j/j_{\rm max} +
  \mu -1}.
\end{equation}
Here $M(<j)$ is the halo mass with specific angular momentum less than
$j$, $\Mvir$ is the halo's virial  mass, $\mu$ is a free parameter, and
$j_{\rm max}$  is the maximum specific  angular momentum in  the halo. 
The value of $j_{\rm max}$ is given by
\begin{equation}
j_{\rm max} = \frac{\Jvir/\Mvir}{(\mu-1)[-\mu\ln(1-\mu^{-1})-1]}, 
\end{equation}
where $\Jvir/\Mvir$  is determined by  the spin parameter.   B01 found
$\mu-1$  to  be log-normally  distributed  with  a  mean and  standard
deviation   in   $\log(\mu-1)$   of  $\simeq-0.6$   and   $\simeq0.4$,
respectively.

Sharma  \& Steinmetz  (2005,  hereafter  SS05) used  a  series of  non
radiative N-body/SPH simulations in  a \LCDM cosmology to study
the growth  of angular momentum  in galaxy systems.  SS05  showed that
Eq.~\ref{eqn:mjbul} is unable to describe the specific angular momentum
distribution  required to make  an exponential  disk in  an NFW  halo. 
They introduced an  alternative function that is able  to describe the
specific angular  momentum distribution of exponential  disks, as well
as those of the gas and dark matter in their simulations.
\begin{equation}
m(j)=\gamma(\alpha,\frac{j}{j_{\rm d}}), \;\; j_{\rm d} = \frac{\Jvir}{\Mvir} \frac{1}{\alpha}
\end{equation}
where  $\gamma$ is  the  incomplete gamma  function.   SS05 found  the
distributions  of  $\alpha$  is  log-normally  distributed  with  mean
$\log\alpha\simeq-0.05$, standard deviation in $\log\alpha\simeq0.11$.

\subsection{Conservation of Angular Momentum and Halo Contraction}
\label{sec:am}
In order to ensure that  specific angular momentum is conserved in our
model,  rather than  keeping track  of  how much  mass is  at a  given
radius, $R$,  we keep track  of how much  mass is at a  given specific
angular momentum,  $j$.  Thus,  under the simplifying  assumption that
$V_{\rm  circ}= [G M(<R)/R]^{1/2}$,  where $M(<R)$  is the  total mass
within a spherical radius, $R$, the radius that corresponds to a given
$j$ is given by
\begin{equation}
R = \frac{j^2}{G M(<j)}.
\end{equation}
This has a number of desirable  properties: 1) At each time step it is
trivial to calculate how much cold gas is added to each bin in $j$. 2)
Over  time,  as  the  potential  well changes,  the  specific  angular
momentum of  the baryons is automatically conserved;  3) The adiabatic
contraction (Blumenthal  \etal 1986)  of the halo  in response  to the
cooling of  the baryons is  automatically taken into account.   4) The
resulting radial grid is adaptive,  as the mapping between $j$ and $R$
depends on the amount of mass enclosed.

As  discussed in  Dutton \etal  (2007)  and DB08,  models with  halo
contraction (and  standard stellar IMF's) are unable  to reproduce the
zero points of the $VMR$ relations as well as the low galaxy formation
efficiency  required to reconcile  the halo  mass function  and galaxy
stellar mass  function.  While there  are processes such  as dynamical
friction  and   impulsive  mass  loss   that  can  expand   the  halo,
implementing these in a galaxy  evolution model is a non-trivial task.
Thus for simplicity we wish to consider a model in which the dark halo
does not respond to galaxy formation. To calculate the mapping between
$j$ and  radius, for  the case of  {\it no adiabatic  contraction}, we
solve the equation
\begin{equation}
  R = \frac{j^2/G}{M_{\rm halo}(R) + M_{\rm disk}(<j)},
\end{equation}
where $M_{\rm halo}(R)$ is the mass (within a spherical radius $R$) of
the dark matter plus hot gas  halo in the absence of galaxy formation,
and $M_{\rm  disk}(<j)$ is the mass  of the disk (gas  plus stars) with
specific angular momentum less than $j$.  This way the self-gravity of
the disk is included but adiabatic contraction is ignored.

\subsection{Star Formation}
Following Blitz \& Rosolowski (2006) we assume that star formation
takes place in dense molecular gas, traced by hydrogen cyanide (HCN),
with a constant star formation efficiency:
\begin{equation}
\label{eq:SKmol}
\frac{\Sigma_{\rm SFR}}{[\Msun \rm  pc^{-2} Gyr ^{-1}]} = \frac{\tilde\epsilon_{\rm SF}}{[\rm Gyr^{-1}]} \frac{\Sigma_{\rm mol, HCN}}{[\Msun \rm pc^{-2}]},
\end{equation}
where $\tilde\epsilon_{\rm SF} \simeq  10-13 \rm {Gyr}^{-1}$ (Gao \& Solomon
2004, Wu \etal 2005).  Expressing  this equation in terms of the total
gas content:
\begin{equation}
\frac{\Sigma_{\rm SFR}}{[\Msun \rm  pc^{-2} Gyr ^{-1}]} = \frac{\tilde\epsilon_{\rm SF}}{[\rm Gyr^{-1}]} \frac{\Sigma_{\rm gas}}{[\Msun \rm pc^{-2}]} \,f_{\rm mol}\, {\cal R}_{\rm HCN},
\end{equation}
where ${\cal R}_{\rm HCN} = \Sigma_{\rm mol, HCN}/\Sigma_{\rm mol}$ is
the ratio between  the dense molecular gas (as traced  by HCN) and the
total molecular gas, and $f_{\rm mol}$ is the molecular gas fraction.

The fraction of  gas that is molecular, $f_{\rm  mol}$, can be defined
in terms  of the  mass ratio between  molecular and atomic  gas, $\cal
R_{\rm mol}$ by
\begin{equation}
\label{eq:fmol}
f_{\rm mol} = \frac{\cal R_{\rm mol}}{{\cal R}_{\rm mol}+1}.
\end{equation}
Empirically Wong \& Blitz (2002)  and Blitz \& Rosolowsky (2004; 2006)
have argued that ${\cal R}_{\rm mol}$ is determined to first order by the mid
plane pressure, $P_{\rm ext}$.  The most recent relation from Blitz \&
Rosolowsky (2006) is
\begin{equation}
\label{eq:Rmol}
{\cal R}_{\rm mol} = \frac{\Sigma_{\rm mol}}{\Sigma_{\rm atom}} 
= \left[ \frac{P_{\rm ext}/k}{4.3\pm0.6 \times 10^4} \right]^{0.92\pm0.1},
\end{equation}
where $k$ is Boltzmann's constant, and $P_{\rm ext}/k$ is in cgs units.
For a gas plus stellar disk the mid plane pressure is given, to
within 10\% by (Elmegreen 1993)
\begin{equation}
\label{eq:Pext}
P_{\rm ext} \simeq \frac{\pi}{2} G \Sigma_{\rm gas} \left [
  \Sigma_{\rm gas} + 
\left(\frac{\sigma_{\rm gas}}{\sigma_{\rm star}}\right) \Sigma_{\rm
  star} \right],
\end{equation}
where  $\sigma_{\rm gas}$  and  $\sigma_{\rm star}$  are the  velocity
dispersions of  the gas and stellar disk,  respectively. For simplicity
we  will assume  $\sigma_{\rm gas}/\sigma_{\rm  star}=0.1$.

Based on the arguments and references in Blitz \& Rosolowski (2006) we
adopt the following relation for ${\cal R}_{\rm HCN}$
\begin{equation}
{\cal R}_{\rm HCN} = 0.1 \left( 1 + \frac{\Sigma_{\rm mol}}{[200 \, \Msun \rm pc^{-2}]} \right)^{0.4}. 
\end{equation}
In the low pressure,  low molecular density regime, ${\cal R}_{\rm HCN}\simeq
0.1$, and thus Eq.~(\ref{eq:SKmol}) asymptotes to
\begin{equation}
\label{eq:SKlow}
\frac{\Sigma_{\rm SFR}}{[\Msun \rm pc^{-2} \rm Gyr^{-1}]} = 
\frac{\tilde\epsilon_{\rm SF}}{[\rm Gyr ^{-1}]}\frac{0.013}{[\Msun\,\rm pc^{-2}]}\, \left(\frac{\Sigma_{\rm gas}}{[\Msun \rm pc^{-2}]}\right)^{2.84}. 
\end{equation}
In the  high pressure, high  molecular density regime,  ${\cal R}_{\rm
  HCN}   \propto  \Sigma_{\rm  mol}^{0.4}$,   and  eq.(\ref{eq:SKmol})
asymptotes to the familiar Schmidt-Kennicutt (SK) relation
\begin{equation}
\label{eq:SKhigh}
\frac{\Sigma_{\rm SFR}}{[\Msun \rm pc^{-2} \rm Gyr^{-1}]}  
= \frac{\tilde \epsilon_{\rm SF}}{[\rm Gyr^{-1}]}\frac{0.012}{[\Msun\,\rm pc^{-2}]}\, \left(\frac{\Sigma_{\rm gas}}{[\Msun \rm pc^{-2}]}\right)^{1.4}.
\end{equation}
Furthermore, with $\tilde\epsilon_{\rm SF}=13 \,\rm Gyr^{-1}$, we recover the
coefficient of $\epsilon_{\rm SF}=0.16$ of the standard SK relation.

We implement the star formation recipe given by Eq.(\ref{eq:SKmol}) as
follows.  At each time step and  annulus in the disk, we calculate the
star formation  rate.  Then we use the  following approximation (valid
for times  steps small compared to  the star formation  time scale) to
calculate the mass of newly formed stars
\begin{equation}
\Delta M_{\rm star}(R) = A(R) \,\Sigma_{\rm SFR}(R,t)\,\Delta t,
\end{equation}
with  $A$ the  area  of the  annulus,  and $\Delta  t$  the time  step
interval.

\subsection{Supernova Feedback}
We assume that SNe drive mass outflows, and that the outflows
move at the local escape velocity of the disk-halo system.  This
choice is motivated by the fact that it maximizes the mass loss from
the the disk-halo system (lower velocity winds will not escape the
halo, and higher velocity winds will carry less mass).

For our  energy driven  wind model following  Dekel \& Silk  (1986) we
assume that  the kinetic energy  of the wind  is equal to  a fraction,
$\epsilon_{\rm EFB}$,  of the kinetic energy produced  by SN. However,
contrary to Dekel  \& Silk (1986) we apply  this energy condition {\it
  locally} in the  disk as a function of  radius, rather than globally
to the whole galaxy.  Thus the mass ejected from radius, $R$, during a
given time step is given by
\begin{equation}
\Delta M_{\rm eject}(R) = \frac{2\,\epsilon_{\rm EFB}\,E_{\rm SN}\,\eta_{\rm SN}}
{V_{\rm esc}^2(R)}\Delta M_{\rm star}(R).
\end{equation}
Here $\Delta M_{\rm  star}(R)$ is the mass in  stars formed at radius,
$R$,   $E_{\rm SN}=10^{51}$    erg   $\simeq   5.0\times    10^{7}   \,\rm
km^2\,s^{-2}\,  \Msun$   is  the  energy  produced  by   one  SN,  and
$\eta_{\rm SN}=8.3\times10^{-3}$  is the number  of SN  per solar  mass of
stars formed (for a Chabrier (2003) IMF).  

Similarly  for our  momentum  driven  wind model  we  assume that  the
momentum of the wind is  equal to a fraction, $\epsilon_{\rm MFB}$, of
the momentum produced  by SN, thus the mass  ejected from radius, $R$,
during a given time step is given by
\begin{equation}
  \Delta M_{\rm eject}(R) = \frac{\epsilon_{\rm MFB}\,p_{\rm SN}\,\eta_{\rm SN}}{V_{\rm esc}(R)}\Delta M_{\rm star}(R).
\end{equation}
Here $p_{\rm SN}=3\times 10^{4}\Msun\kms$  is the momentum produced by
one SN, assuming  that each SN produces $\simeq  10 \Msun$ of material
moving  at $v\simeq 3000  \kms$ (Murray,  Quataert \&  Thompson 2005).
Note  that this  corresponds to  a kinetic  energy of  $4.5\times 10^7
\,\rm \Msun \, km^{2}\,s^{-2}\simeq 10^{51}$ erg.

We assume that the ejected mass is lost forever from the system: the
ejected mass is not considered for later infall, and the corresponding
metals are not used to enrich the infalling gas. This is clearly a
dramatic oversimplification, but we make this choice to maximize the
amount of gas that is lost from the system.

\subsection{Overview of Parameters}
The input parameters of our models are as follows.

(1) Cosmology: $\Omega_{\rm m,0}, \Omega_{\Lambda}$, $\Omega_{\rm b}$,
$\sigma_8$, $h$, $n$.   In this paper we adopt  a flat \LCDM cosmology
motivated  by the  5th year  WMAP results  (Dunkley \etal  2009), with
$\Omega_{\rm m,0}=0.258, \Omega_{\Lambda}=0.742, \Omega_{\rm b}=0.044,
\sigma_8=0.80, h=0.7$, and $n=1$.
    
(2) Halo structure: $K, F, \sigma_{\ln c}$. We adopt the Bullock \etal
(2001a) model with $F=0.01$, $K=3.7$, and $\sigma_{\ln c}=0.25$. These
parameters  reproduce  the  distribution  of  halo  concentrations  of
relaxed  dark  matter  haloes  in cosmological  N-body  simulations
(e.g. Wechsler \etal 2002; Macci\'o \etal 2007; 2008)

(3)  Angular   momentum  distribution:  $\bar{\lambda}$,  $\sigma_{\ln
  \lambda}$,   $\alpha$    (or   $\mu$),   $\sigma_{\log\alpha}$   (or
$\sigma_{\log(\mu-1)}$).  As  fiducial values  we adopt a  median spin
parameter   $\bar{\lambda}=0.025$    with   a   scatter   $\sigma_{\ln
  \lambda}=0.35$ (see DB08).  For the angular
momentum distribution  we consider two fitting  functions: the Bullock
\etal  (2001b)   formula  with  median   ${\mu-1}=0.3$,  $\sigma_{\log
  (\mu-1)}=0.45$; and  the Sharma \&  Steinmetz (2005) formula  with a
median $\alpha=0.9$  and $\sigma_{\log\alpha}=0.11$. For  our fiducial
model  we  adopt  the  Sharma  \& Steinmetz  (2005)  angular  momentum
profile.

(4)  Star  formation:  $\tilde  \epsilon_{\rm  SF}$.  We  use  a  star
formation  model  based  on  dense  molecular  gas,  which  we  compute
following    Blitz   \&    Rosolowsky    (2006). We adopt $\tilde\epsilon_{\rm
  SF}=13 \rm \,Gyr^{-1}$.

(5) Feedback: $\epsilon_{\rm EFB}$, $\epsilon_{\rm MFB}$, $\eta_{\rm
  SN}$, $E_{\rm SN}$, $p_{\rm SN}$. We adopt an energy per SN of
$E_{\rm SN}=10^{51}$ erg, a momentum per SN of $p_{\rm SN}=3\times
10^{4} \Msun\kms$, a SN rate of $\eta_{\rm SN}=4\times10^{-3}$ per
solar mass of stars formed.  We treat $\epsilon_{\rm EFB}$ and
$\epsilon_{\rm MFB}$ as free parameters, with fiducial values of
$\epsilon_{\rm EFB}=0.25, \epsilon_{\rm MFB}=0$.
    
(6) Stellar populations and chemical  enrichment: ${\cal R}, y, Z_{\rm
  hot}$, and the choice of  initial mass function (IMF).  We adopt the
Chabrier IMF, a  return fraction ${\cal R}=0.35$, and  a stellar yield
$y=0.02$, and a hot gas metallicity of $0.002$ (i.e. $0.1Z_{\odot}$).


\section{Surface Density Profiles}
\label{sec:exp}

The  surface density  profile  of a  centrifugally  supported disk  is
determined by two things:
\begin{enumerate}
\item the distribution of specific angular momentum (AMD), $j$, of the
  baryons (i.e. stars and cold gas); and
\item the shape and amplitude of the rotation curve.
\end{enumerate}
The AMD of the baryons may be modified from that produced by tidal
torques, by inefficient cooling and SN driven outflows. The difference
between the surface density profile of the stellar disk and baryonic
disk is then determined by the (radially dependent) efficiency at
which the cold gas is turned into stars.  Another complication is that
observed light is a biased tracer of the stellar (and baryonic) mass.
Below we discuss each of these effects for a fiducial model galaxy.
This model has a total virial mass of $\Mvir=9.0 \times 10^{11}\Msun$,
a median halo concentration of this mass, and unless other-wise
specified a halo spin parameter of $\lambda=0.025$.  This model also
has no halo contraction, as models with halo contraction are unable to
reproduce the zero points of the Tully-Fisher (TF) (Tully \& Fisher
1977) and size-mass relations (DB08). However, as we show in \S\S
\ref{sec:ac} \& \ref{sec:exp} none of our main results are sensitive
to this assumption.

\begin{figure*}
\begin{center}
\centerline
{\psfig{figure=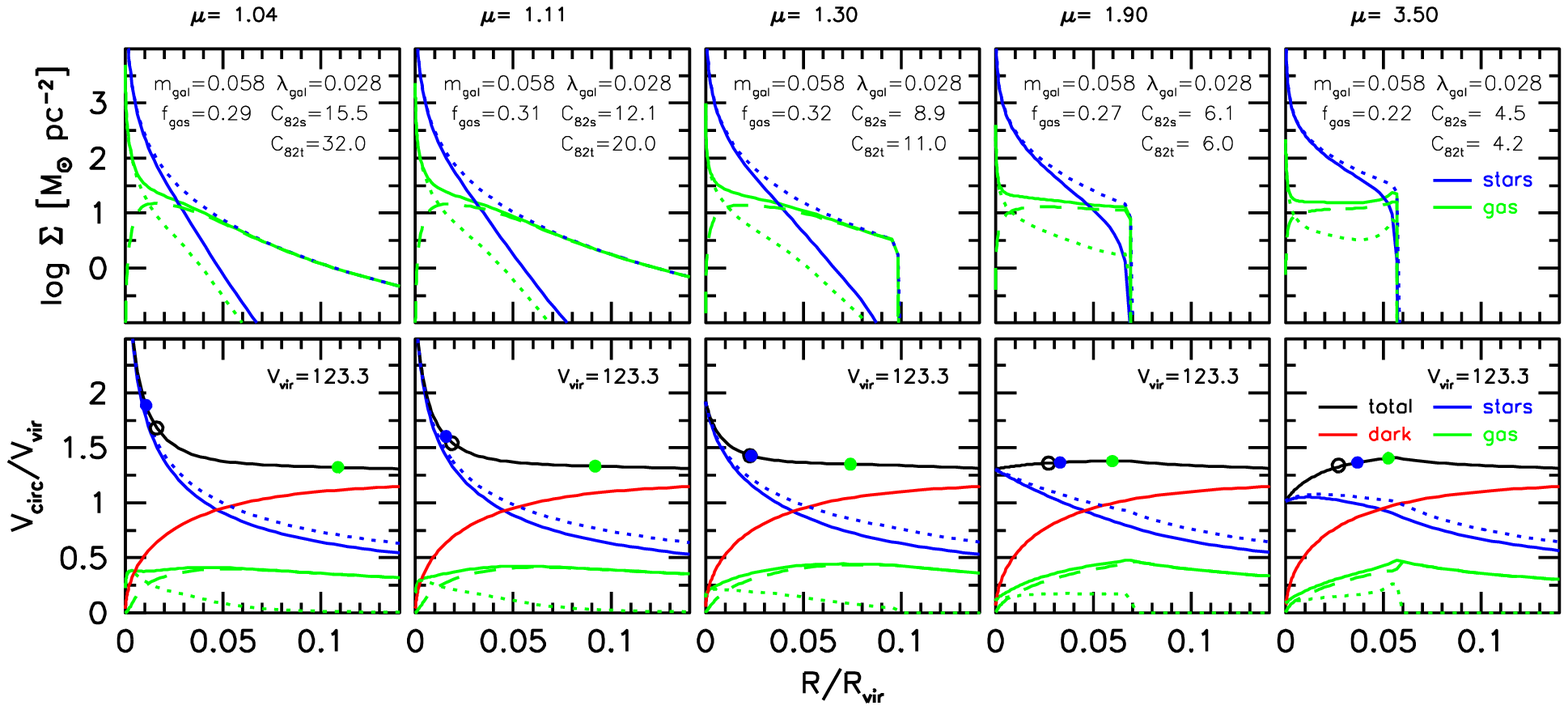,width=0.80\textwidth}}
\centerline
{\psfig{figure=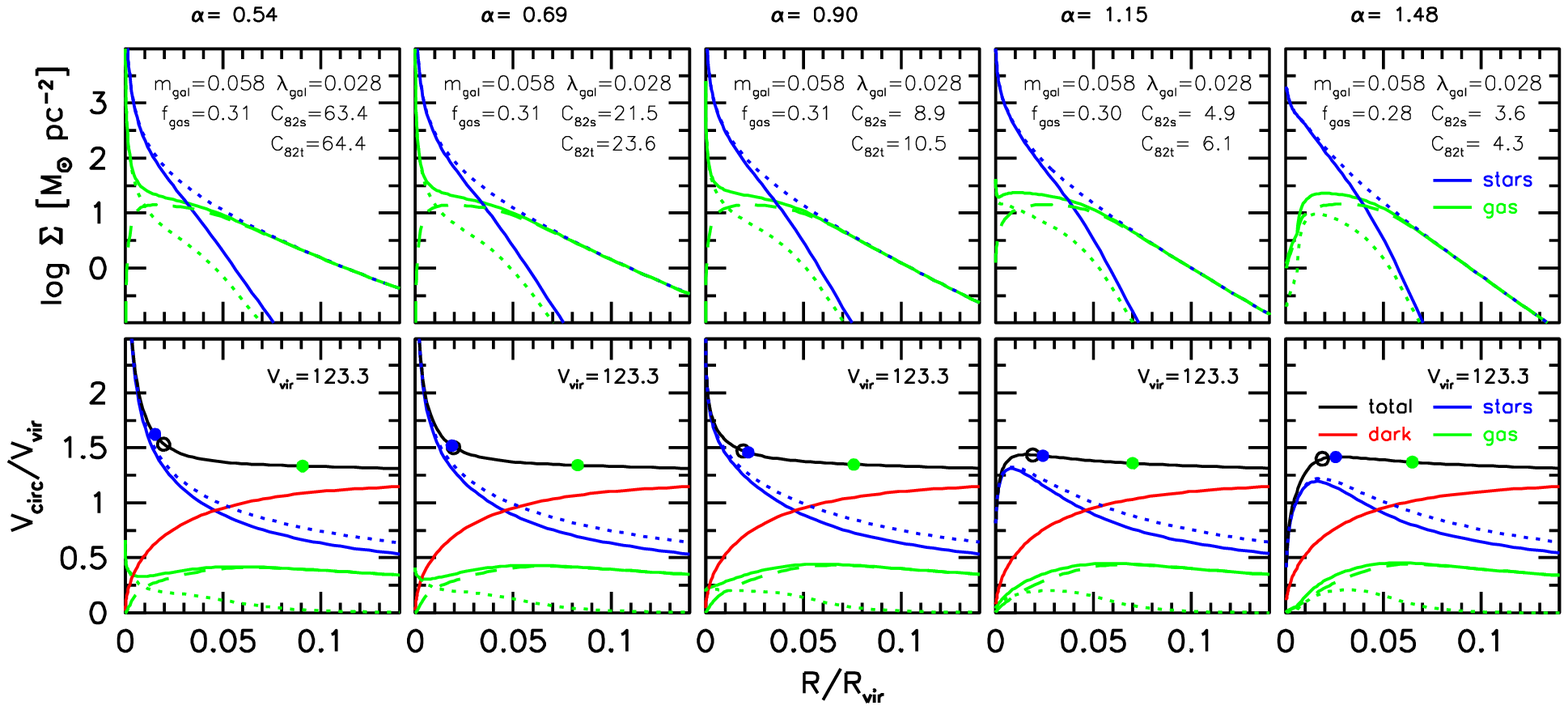,width=0.80\textwidth}}
\caption[]{\footnotesize Effect of angular momentum shape parameter on
  surface density and circular velocity profiles. Radii are expressed
  in terms of the virial radius, $\Rvir$, and velocities in terms of
  the circular velocity at the virial radius, $\Vvir$.  For the
  cosmology we adopt, the virial radius (in units of
  $h_{70}^{-1}\kpc$) is $\simeq 2.068$ times the virial velocity (in
  units of $\kms$).  All models have $\mgal=0.058$, $\lambda=0.028$,
  and $\Mvir = 10^{11.8} \hMsun$. The upper panels use the B01 angular
  momentum profile (specified by $\mu$), while the lower panels use
  the SS05 angular momentum profile (specified by $\alpha$).  The five
  values of the angular momentum shape parameter correspond to the
  mean, $\pm 1\sigma$ and $\pm2\sigma$ of the distributions found in
  cosmological simulations.  The stellar disk is given by the blue
  solid lines, the total disk by the blue dotted lines.  The gas disk
  is given the the solid green lines, the molecular gas is given by
  the dotted green lines, and the atomic gas is given by the dashed
  green lines.  The circular velocity of the halo is given by the
  solid red line and the total circular velocity is given by the solid
  black line.  The circles show the circular velocity at three
  characteristic radii: 2.15 stellar disk scale lengths (black open
  circles); the radius enclosing 80\% of the stellar mass, $R_{80 \rm
    s}$ (blue filled circles); and the radius enclosing 80\% of the
  cold gas mass, $R_{80 \rm c}$ (green filled circles).  For each model
  the cold gas fraction, $f_{\rm gas}$, and stellar disk concentration,
  $C_{82\rm s}$, and the baryonic disk concentration, $C_{82\rm t}$, are given.}
\label{fig:sv-BS}
\end{center}
\end{figure*}

\subsection{The Role of the Initial Angular Momentum Distribution}
In the standard picture of disk galaxy formation the distribution of
$j$ of the cooled gas is the same as that of the hot gas and dark
matter.  As discussed in \S~\ref{sec:models}, N-body simulations have
shown that there is a universal $j$ profile, described by one
parameter ($\mu$ in B01; $\alpha$ in SS05).  In Fig.\ref{fig:sv-BS} we
show the effect of the initial AMD shape parameter on the surface
density and circular velocity profiles for our fiducial model galaxy.
The upper panels use the B01 AMD with $\mu=1.04, 1.11, 1.30, 1.90$,
and $3.50$, while the lower panels use the SS05 AMD with $\alpha=0.54,
0.69, 0.90, 1.15$, and $1.48$.  For both AMD's these values correspond
to the $-2,-1,0,1$, and $2\sigma$ of the distribution found in
cosmological simulations.

In  order to  isolate  the effects  of  the AMD  from  the effects  of
feedback, we fix the galaxy mass fraction and galaxy spin parameter to
be equal to the values for  a galaxy with $\alpha=0.9$ and our favored
energy    driven   feedback    model    ($\epsilon_{\rm   EFB}=0.25$):
$\mgal=0.058$ and $\lamgal=\lam=0.028$.

Both the B01 and SS05 AMD's result in roughly the same surface density
profiles and rotation  curve shapes.  The generic result  is for lower
values of $\mu$ and $\alpha$ to result in higher surface densities and
circular velocities  at small radii.   As $\mu$ and  $\alpha$ increase
the  central densities  decrease  and the  circular velocity  profiles
become flatter.   There are however two notable  differences.  The B01
profile  has a  maximum  $j$, which  results  in a  truncation in  the
surface density  profile, this truncation  occurs at a  smaller radius
for larger  values of  $\mu$.  The SS05  profile has less  low angular
momentum material  for high  values of $\alpha$  than the  B01 angular
momentum profile for high values of $\mu$. 

SS05 showed  that the AMD at large  $j$ depends on the  method used to
calculate the AMD.   In particular, the cell method  used by B01 tends
to result  in a truncation  of the AMD  at large $j$, which  is better
fitted by the B01 profile. The particle method used by SS05 results in
a smooth and extended AMD at  large $j$, which is better fitted by the
SS05 profile.  If  one removes the outer 5\%  of the $j$-distribution,
SS05 showed that  both profiles provide equally good  fits to the AMDs
computed using the cell and particle methods.

The concentration parameters for each model are given in
Fig.~\ref{fig:sv-BS}.  The concentration parameter of the stellar
($C_{82\rm s}$) and baryonic disks ($C_{82\rm t}$) are defined as the
ratio between the radii enclosing 80 and 20\% of the disk
stellar/baryonic mass.  For reference an exponential profile (S\'ersic
$n=1$) corresponds to a concentration of $C_{82}=3.6$, a S\'ersic
$n=2$ profile corresponds to a $C_{82}=5.8$, and a de Vaucouluers
profile (S\'ersic $n=4$) corresponds to a $C_{82}=11.5$.  

The B01 AMD results in less dispersion in the concentrations of the
stellar and baryonic disks than the SS05 AMD, both at high and low
values of $\alpha$ or $\mu$.  It should be noted that some of the very
high concentrations $C_{82\rm s} \simeq 100$ produced with the SS05
AMD are unrealistic.  These are caused by the models having too much
mass with low specific angular momentum, and our requirement that this
specific angular momentum be conserved.  In order for mass with low
specific angular momentum to be in centrifugal equilibrium it has to
be at small radii. However, when the baryons dominate the potential at
small radii, this increases the circular velocity, which in turn
requires the baryons move to even small radii, and so on.  When
centrifugal equilibrium is reached, the inner disk is very compact.
In practice the disk will become unstable before it reaches
centrifugal equilibrium, and the baryons in the inner part of the
galaxy will become at least partially supported by random motions,
which will result in lower concentrations.

\begin{figure*}
\centerline
{\psfig{figure=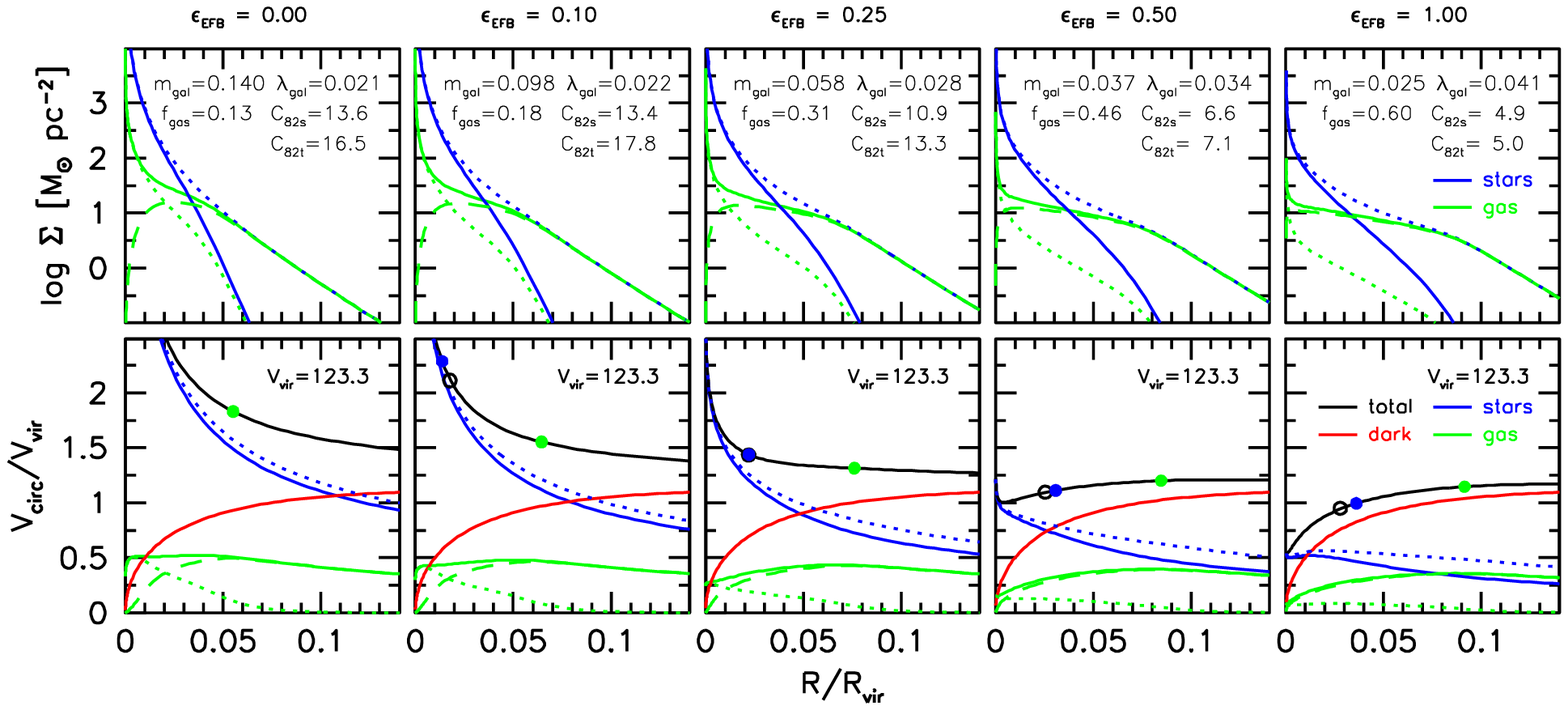,width=0.80\textwidth}}
\centerline
{\psfig{figure=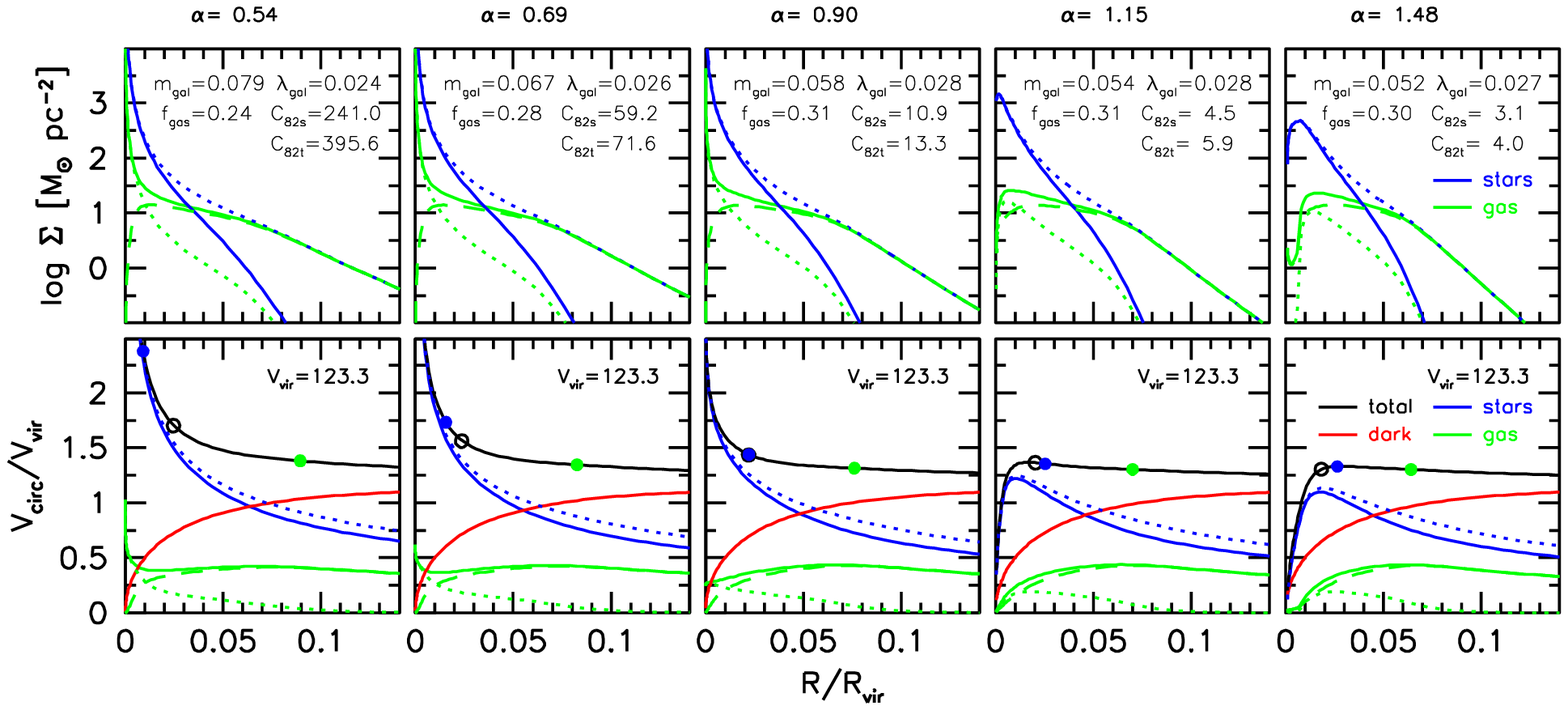,width=0.80\textwidth}}
\caption[]{\footnotesize Effect of feedback on the surface density and
  rotation curves using the SS05 angular momentum profile.  The upper
  panels have $\alpha=0.9$ with $\epsilon_{\rm EFB}$ varying from 0 to
  1, while the lower panels have $\epsilon_{\rm EFB}=0.25$ with
  $\alpha$ varying over the $2\sigma$ range. The line types and colors
  are the same as in Fig.\ref{fig:sv-BS}.  Without feedback (upper
  panels, far left) the model has too many baryons, which results in a
  disk that is too small, too concentrated, gas poor, and which
  rotates too fast. As the efficiency of feedback in increase (upper
  panels, from left to right), the baryon fraction decreases, which
  results in a larger disk, with lower concentration, higher gas
  fraction and lower rotation velocity. A model with maximal feedback
  (upper panels, far right) has a large, low concentration, gas rich
  disk, with a rising inner rotation curve. }
\label{fig:sv-S-FB}
\end{figure*}

\subsection{The Role of Star Formation}
Fig.~\ref{fig:sv-BS} also shows that the surface density profile of
the stellar disk is very different than that of the total baryonic
disk.  The stellar disks are closer to exponential and have smaller
sizes than the baryonic disks.  These differences are due to the
increased efficiency of star formation at higher gas densities (the
Schmidt-Kennicutt law).  More specifically, in the star formation
model that we adopt, stars only form out of molecular gas, and the
molecular fraction is a strong function of the surface density of
stars and gas.  In Fig.~\ref{fig:sv-BS} the surface density of the
molecular gas is given by the dotted green lines, and that for the
atomic gas by green dashed lines. At small radii the gas is mostly
molecular, whereas at large radii the gas is mostly atomic, and hence
results in very inefficient star formation at low gas densities.

Our models thus predict that gas disks should have larger sizes than
stellar disks, and that the baryonic disks are in general not
exponential.  Thus a study of the baryonic surface density profiles of
galaxy disks would provide useful constraints to the initial AMDs and
the efficiency of star formation at low gas densities.

An interesting consequence of the inefficiency of star formation at
low gas densities, is that this results in a outer disk with a shorter
scale length than the main disk. Such ``disk breaks'' are often seen
in observed spiral galaxies (e.g. Pohlen \& Trujillo 2006). Why some
galaxies show breaks, and others are purely exponential to large radii,
and still others show anti-truncations (i.e. larger outer disk scale
lengths), is an open question.  If the efficiency of star formation at
low gas densities varies between galaxies this could provide an
explanation for the wide range in outer disk profiles observed in
spiral galaxies. A study of baryonic disk density profiles would thus
provide a means of distinguishing between a star formation efficiency
origin vs a dynamical origin (e.g. Debattista \etal 2006; Younger
\etal 2007; Ro{\v s}kar \etal 2008; Foyle, Courteau \& Thacker 2008)
for these features.

\begin{figure*}
\begin{center}
\centerline
{\psfig{figure=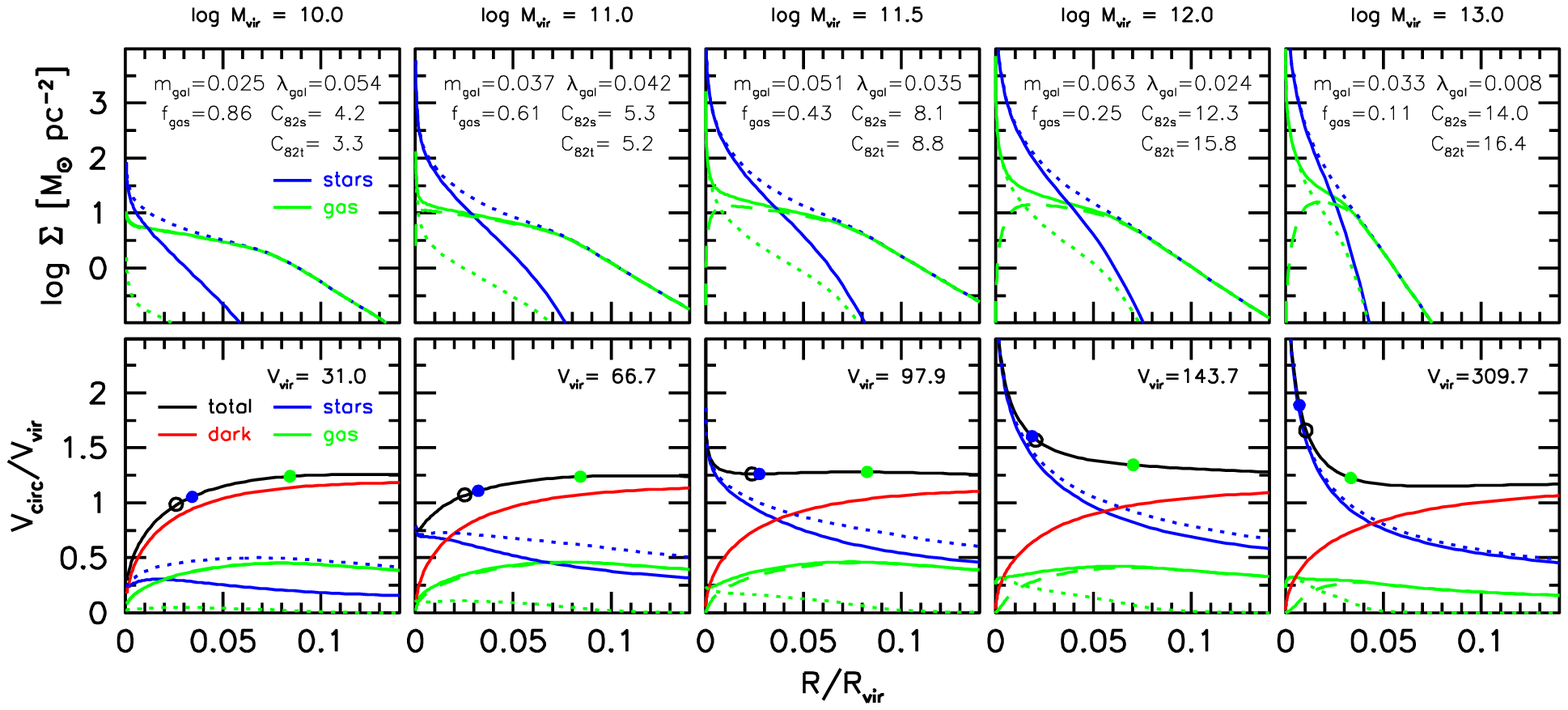,width=0.80\textwidth}}
\centerline
{\psfig{figure=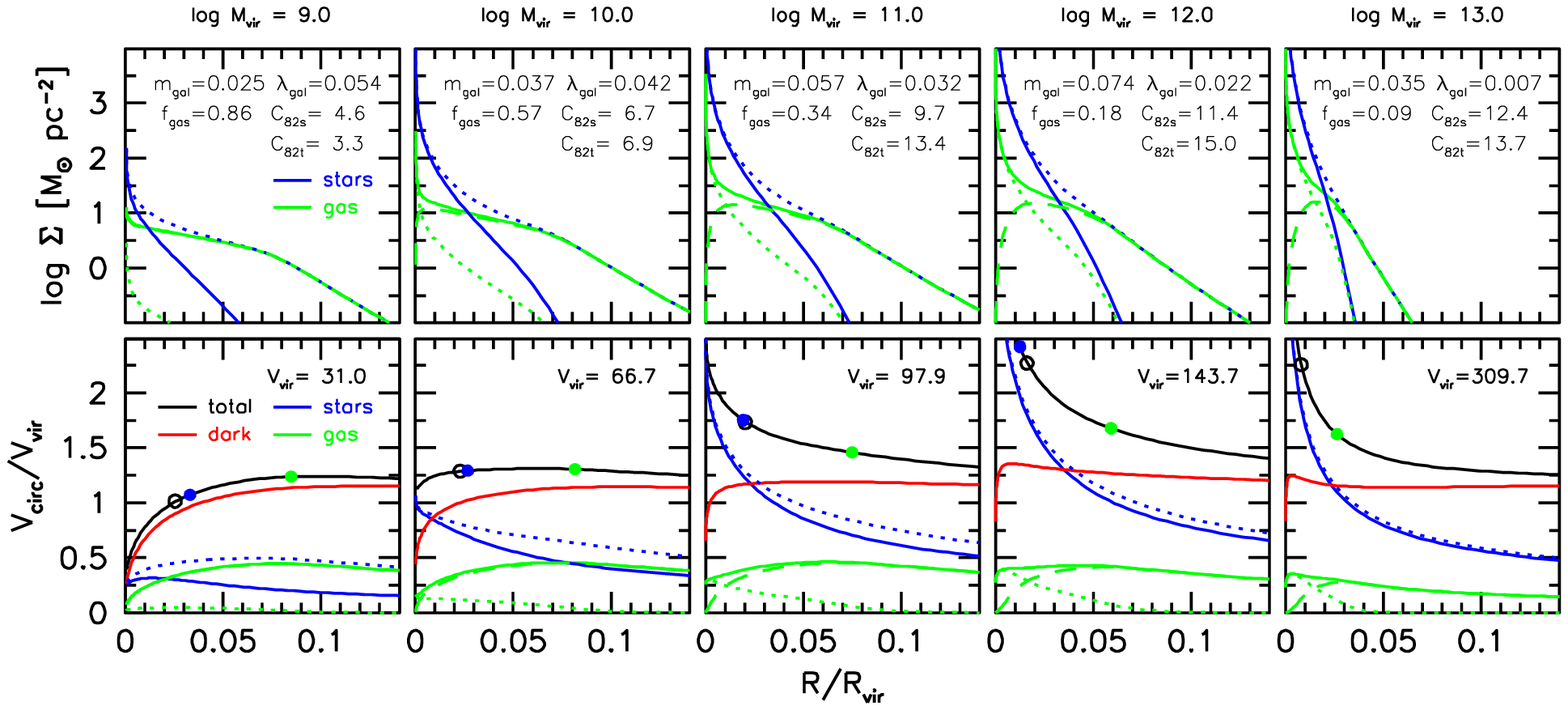,width=0.80\textwidth}}
\caption[]{\footnotesize Effect of halo mass on surface density and
  circular velocity profiles in a model with feedback.  All models
  have $\epsilon_{\rm EFB}=0.25$, $\alpha=0.90$ and
  $\lambda=0.025$. The redshift zero virial mass (in $h=1$ units) is
  given above each column, the corresponding virial velocity is given
  in the velocity panels. The upper panels are for a model without
  adiabatic contraction, the lower panels include adiabatic
  contraction. The line types and colors are as in
  Fig.~\ref{fig:sv-BS}.  Low mass haloes result in low galaxy mass
  fractions ($\mgal$), high galaxy spin parameters ($\lamgal$), high
  gas fractions ($f_{\rm gas}$), low stellar concentrations ($C_{82\rm
    s}$), and dark matter dominated inner rotation curves.  As halo
  mass increases $\mgal$ increases (up until $\Mvir\simeq
  10^{12}\hMsun$), $\lamgal$ decreases, $f_{\rm gas}$ decreases and
  $C_{82\rm s}$ increases, and the inner rotation curve becomes
  dominated by baryons. Adiabatic contraction increases the density of
  the halo, especially in the inner few percent of the virial
  radius. The differences between models with and without adiabatic
  contraction are in general small, with the exception of the
  amplitude of the inner rotation curve. See text for further details.}
\label{fig:sv-mvir}
\end{center}
\end{figure*}

\subsection{The Role of Feedback}
The models  in Fig.~\ref{fig:sv-BS} had  the galaxy mass  fraction and
spin  parameter fixed. However,  in a  halo with  $\Mvir\simeq 9\times
10^{11}\Msun$,  the  cooling efficiency  is  high,  such  that in  the
absence of  feedback the  galaxy mass fraction  is $\simeq  80\%$.  In
order to  produce galaxies with realistic mass  fractions $\lta 33\%$,
mass outflows (i.e. feedback) are required.

The effect  of feedback  on the surface  density profiles is  shown in
Fig.\ref{fig:sv-S-FB}. The upper panels show a model with the SS05 AMD
and $\alpha=0.90$  and with the feedback efficiency  varying from left
to right: $\epsilon_{\rm EFB}=0.00, 0.10, 0.25, 0.50$, and $1.00$. The
lower panels show a model with $\epsilon_{\rm EFB}=0.25$, and the SS05
AMD varied  over the $2\sigma$ range as  in Fig.~\ref{fig:sv-BS} (this
corresponds to model III in DB08).

Feedback effects the surface density profile in two ways: 1) it
removes mass, decreasing the galaxy mass fraction, and 2) it
preferentially removes low angular momentum material which also modifies
the AMD.  The galaxy mass fractions, $\mgal=\Mgal/\Mvir$ (defined as
the ratio between the galaxy mass and the total mass of the
galaxy-halo system), and galaxy spin parameters,
$\lamgal=(\jgal/\mgal)\lambda$, (where $\jgal=\Jgal/\Jvir$ is the
ratio between the total angular momentum of the galaxy and that of the
galaxy-halo system) are given for each model.  As the feedback
efficiency is increased the galaxy mass fraction decreases and the
galaxy spin parameter increases (upper panels of
Fig.~\ref{fig:sv-S-FB}).  Both of these result in a more extended disk
and a more gradually rising rotation curve.  By decreasing the
amplitude of the rotation curve at small radii gas will settle in
centrifugal equilibrium at a larger radius, hence resulting in a less
concentrated disk.  Feedback also preferentially removes low angular
momentum material, which also directly results in a less concentrated
disk.

In  these  models the  halo  was  uncontracted  (as models  with  halo
contraction  over-predict the zero  point of  the TF  relation).  Halo
contraction would  make the trends with  feedback efficiency stronger,
as high  galaxy mass fractions  will result in more  halo contraction,
higher circular velocities and hence higher stellar central densities.

The amount of gas ejected as  a function of radius depends on both the
star formation rate  and the depth of the  potential well.  Since both
star formation is  more efficient and the potential  well is deeper in
the  centers  of  galaxies  it  is not  obvious  which  effect  should
dominate.   Indeed, as  shown in  DB08, globally,  the depth  of the
potential well dominates over the star formation efficiency, such that
galaxies  in less  massive haloes  eject  a higher  fraction of  their
accreted gas.  However, locally, for  a fixed halo mass, smaller (i.e.
lower spin  parameter) galaxies are  more efficient at  removing their
baryons, because the higher star  formation rates are more than enough
to compensate for the deeper potential well.

The  lower  panels  of   Fig.~\ref{fig:sv-S-FB}  show  the  effect  of
variation  in the AMD  shape parameter,  $\alpha$, over  the $2\sigma$
range  as found  in  cosmological simulations.   These  models can  be
compared with those in the lower panels of Fig.~\ref{fig:sv-BS}, where
the  galaxy   mass  fractions  and  spin  parameters   were  fixed  at
$\mgal=0.058$      and      $\lamgal=0.028$     respectively.       In
Fig.~\ref{fig:sv-S-FB} the  galaxy mass fractions  and spin parameters
are  determined   by  the   efficiencies  of  feedback   and  cooling.
Qualitatively  the  surface  density  profiles and  circular  velocity
profiles are similar  between the two models. However,  there are some
small differences  which highlight the impact the  distribution of the
baryons have  in determining the amount  of mass that  can be removed.
Larger values of $\alpha$ result  in lower galaxy mass fractions (i.e.
feedback is more efficient). This  is a direct result of the shallower
potential  well created by  a less  concentrated baryonic  disk, which
makes it  easier to eject baryons  from the center of  the galaxy. For
the    model     with    $\alpha=1.48$,    a     comparison    between
Fig.~\ref{fig:sv-BS}  and Fig.~\ref{fig:sv-S-FB}  shows  that feedback
has removed  the ``bulge''  component, creating a  hollow core  in the
disk density profile. This results in a stellar disk less concentrated
than  exponential.  On  the other  hand, for  low values  of $\alpha$,
there is more low angular momentum material, which results in a deeper
potential well, from which it is harder to remove gas.

\begin{figure*}
\begin{center}
\centerline
{\psfig{figure=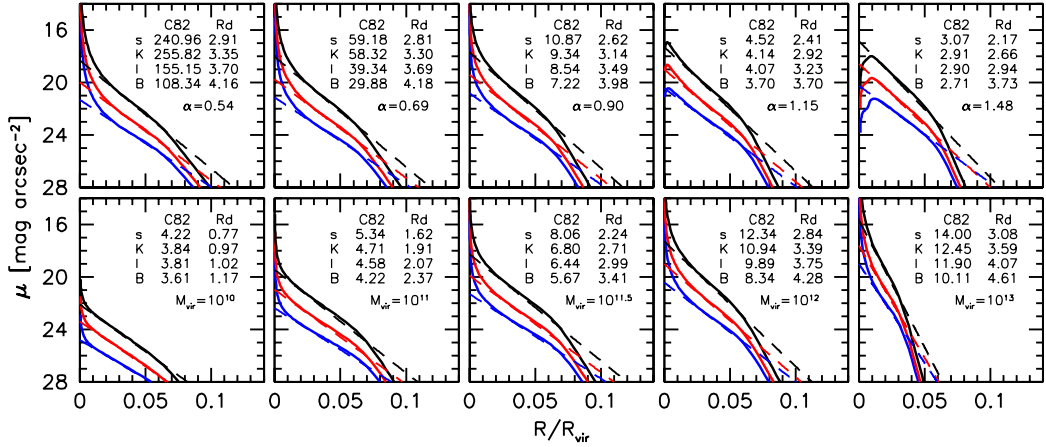,width=0.80\textwidth}}
\caption[]{\footnotesize Effect of stellar populations on
  concentration index and disk scale lengths for model galaxies in the
  lower panels of Fig.~\ref{fig:sv-S-FB} and upper panels of
  Fig.~\ref{fig:sv-mvir}.  For each model the surface brightness
  profiles in B-band (blue), I-band (red) and K-band (black) light are
  shown. The dashed lines show the exponential disk fits.  The
  concentration parameters (defined as the ratio between radii
  enclosing 80 and 20\% of the light (or stellar mass) and disk
  scalelengths are given for each model.}
\label{fig:sb}
\end{center}
\end{figure*}

\subsection{The Role of Halo Mass}
Having discussed the effects of the AMD and feedback efficiency on the
surface  density  profiles  for   a  single  halo  mass,  with  median
concentration and spin  parameters, we now turn our  attention to halo
mass    dependence    of     galaxy    surface    density    profiles.
Fig.~\ref{fig:sv-mvir} shows the surface density and circular velocity
profiles for 5 model galaxies with $\log (\Mvir/[h^{-1}\Msun]) = 10.0,
11.0, 11.5, 12.0,$ and $13.0$. All  of these models have the same halo
angular momentum parameters: $\alpha=0.90$, $\lambda=0.025$, yet their
resulting   galaxy    spin   parameters,   $\lamgal$,    and   stellar
concentrations,   $C_{82\rm  s}$,   vary   systematically  with   halo
mass. Below a  halo mass of $\Mvir = 10^{12} \Msun$  almost all of the
baryons  that accrete  onto the  halo have  had time  to cool,  so the
differences between  the structural properties  of the galaxies  are a
result of the efficiency at which feedback can remove mass and angular
momentum from  the disk and  halo.  As shown  in DB08, mass  loss is
more efficient from lower mass  haloes, and thus feedback has a larger
impact  on the  density profiles  of  low mass  galaxies.  Since  star
formation is biased  to small radii (due to  the density dependence of
the star  formation rate), the mass  that gets removed  has lower than
average  specific  angular  momentum.   The ejection  of  low  angular
momentum material directly lowers  the concentration of the stars. The
reduced  galaxy  mass fractions  and  higher  galaxy specific  angular
momentum  result in  a more  dark matter  dominated  circular velocity
profile.

In   massive  haloes,   $\Mvir  \simeq   10^{13}\hMsun$   feedback  is
inefficient  at removing  mass.   Cooling is  also inefficient,  which
results  in a low  galaxy mass  fraction, $\mgal=0.033$.   But because
cooling occurs  from the inside  out, the resulting spin  parameter of
the galaxy is very  small, $\lamgal=0.008$.  Thus the resulting galaxy
is very compact and  concentrated, which resembles early type galaxies
more than late type galaxies. However, we do not consider these models
to  be realistic,  as the  effects  of mergers  and secular  evolution
(which are not taken into account  in our models) are likely to play a
significant role in producing bulges in high mass haloes.

\subsection{The Role of Adiabatic Contraction}
\label{sec:ac}
The lower panels of Fig.~\ref{fig:sv-mvir} show the same models as the
upper panels but including the effect of adiabatic contraction.
Adiabatic contraction increases the density of the dark matter halo
(red lines in Fig.~\ref{fig:sv-mvir}) in the inner 10\% of the virial
radius. A deeper potential well results in a higher escape velocity,
and thus mass ejection is less efficient. As shown in
Fig.~\ref{fig:sv-mvir} and as discussed in Dutton \& van den Bosch
(2008), models with adiabatic contraction have higher rotation
velocities and smaller sizes than models without adiabatic
contraction. As might be expected the effect on the density profile of
the disk is to increase its concentration, but the effects are in
general small, especially in low mass haloes. Thus adiabatic
contraction does not play a significant role in determining the
density profiles of galaxy disks.

\subsection{The Role of Stellar Populations}
\label{sec:stelpop}
Fig.~\ref{fig:sb} shows surface brightness profiles in K-, I- and
B-bands of the models in the lower panels of Fig.~\ref{fig:sv-S-FB}
(varying AMD shape) and the upper panels of Fig.~\ref{fig:sv-mvir}
(varying halo mass).  In each panel the concentration parameter,
$C_{82}$, in K-, I-, and B-band light as well as in stellar mass is
given.  Concentrations are largest in stellar mass and smallest in
B-band light.  The differences depend both on the concentration of the
stars and the mass of the galaxy, with differences typically a factor
of $\simeq 1.4$.  These differences are caused by color gradients:
inner disks being redder than outer disks.  The color gradients are a
reflection of the inside-out formation of stellar disks, both because
our star formation law results in more efficient star formation at
higher gas densities (which occur at smaller radii), and because the
gas disk is growing with time which reflects the growth of the dark
matter halo.  It is interesting to note that there are non-negligible
differences between concentrations in K-band light and stellar mass,
which we return to in \S\S~\ref{sec:colser} \& \ref{sec:colser2}.
 
Fig.~\ref{fig:sb}  also  shows the  exponential  disk  fits as  dashed
lines.  The  scale lengths in stellar  mass, K-, I-,  and B-band light
are  given for each  model.  The  scale lengths  in B-band  lights are
typically 1.14 times larger than those in I-band light, and 1.25 times
larger than  those in K-band light.  Observationally,  for nearby disk
galaxies,  MacArthur \etal  (2003) finds  a  mean ratio  of B-band  to
R-band  scale  lengths  of  $R_{\rm  d,B}/R_{\rm  d,R}=1.15$,  with  a
standard  deviation of  0.15, and  B-band to  H-band scale  lengths of
$R_{\rm d,B}/R_{\rm d,H}=1.34$ with  a standard deviation of 0.20. Our
models are  consistent with these  observations.  If we  also consider
that extinction  will most likely  cause the observed gradients  to be
over-estimated, this will bring  the models into even better agreement
with the observations.

K-band light  is generally considered  a good tracer of  stellar mass,
but in  our models the scale  length of the K-band  light is typically
1.2 times  larger than  the scale length  in stellar mass.   This also
means that the  scale lengths in B-band light  are typically 1.5 times
larger than  those in  stellar mass. Thus  in order to  derive stellar
mass  scale  lengths  (which  are  a  fundamental  parameter  of  disk
galaxies) from observations, it is  necessary to take into account the
color  gradients  which will  result  in  stellar mass-to-light  ratio
gradients, even in the near IR.

\begin{figure*}
\centerline
{\psfig{figure=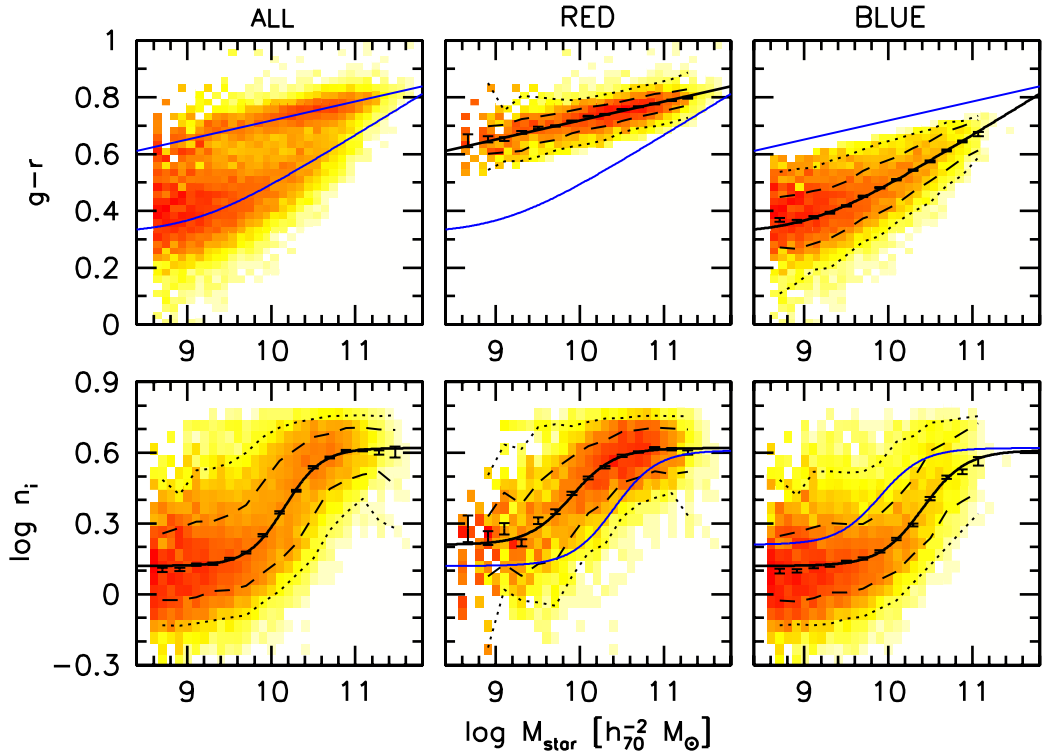,width=0.80\textwidth}}
\caption[    ]{\footnotesize   Color-stellar    mass    and   S\'erisc
  index-stellar  mass  relations  for  galaxies in  the  low  redshift
  NYU-VAGC.  The  color coding corresponds  to the relative  number of
  galaxies in each bin, weighted  by $1/V_{\rm max}$, on a logarithmic
  scale.   Galaxies  are  selected  to  have stellar  mass  $\Mstar  >
  10^{8.6}\Msun$, redshift $0.005 < z <  0.05$, and axis ratio $ b/a >
  0.5$.   We separate  galaxies  into red  (middle)  and blue  (right)
  according  to  their position  in  the  $u-r$  color-$r$  magnitude
  diagram, using the relation in  Baldry \etal (2004).  In addition we
  require that  red galaxies are not  bluer than 0.1 dex  from the red
  sequence, and  that blue galaxies are  at least 0.05  dex bluer than
  the red  sequence.  We  show the median  (solid), 15.9th  and 84.1th
  (dashed), 2.3th and 97.7th (dotted) percentiles of the distributions
  of color and  S\'ersic index in stellar mass bins  of width 0.2 dex.
  The  error bars show  the Poisson  error on  the median.   The thick
  solid  lines   show  fits  to   the  median  relations,   see  Table
  \ref{tab:params} for  fit parameters.  For the red  galaxies we show
  the relations  for the blue galaxies  as thin solid  lines, and vice
  versa for the blue galaxies.}
\label{fig:cn}
\end{figure*}

\begin{figure*}
\begin{center}
\centerline
{\psfig{figure=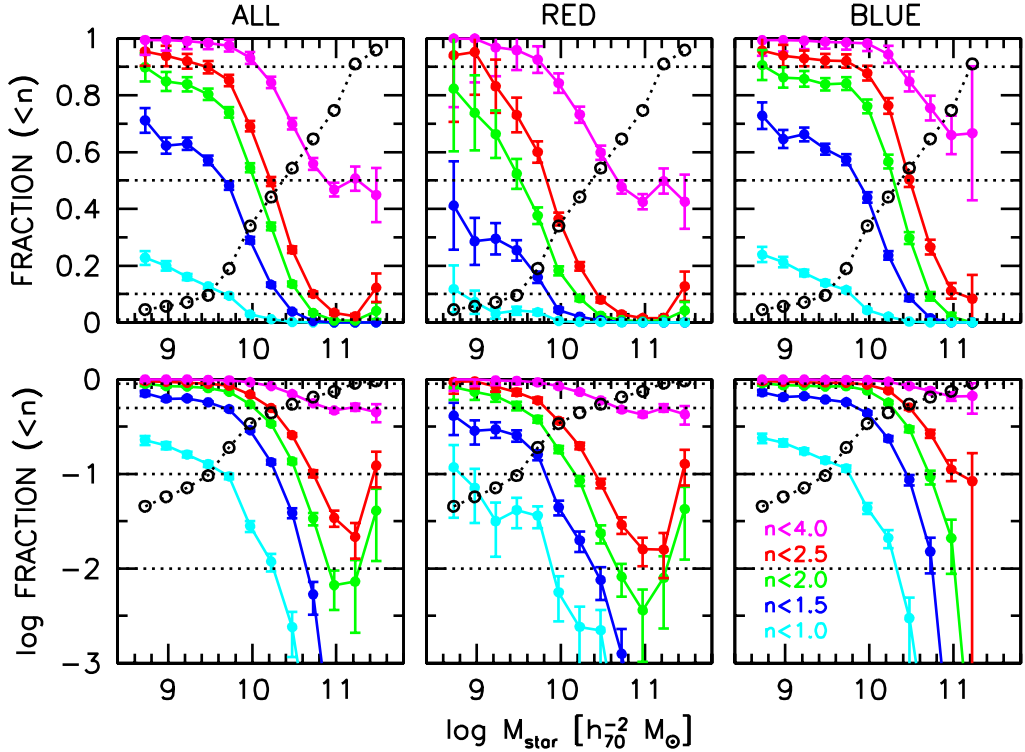,width=0.80\textwidth}}
\caption[  ] {\footnotesize Fraction  of galaxies  at a  given stellar
  mass  with S\'ersic  index less  than  1.0 (cyan),  1.5 (blue),  2.0
  (green),  2.5  (red), and  4.0  (magenta).   The  left panels  show
  results for  all galaxies, middle  panels for red galaxies,  and the
  right panels for  blue galaxies. The black points  show the fraction
  of red galaxies  at a given stellar mass. The  upper panels show the
  fractions in linear  space.  The lower panels show  the fractions in
  log-space,  which  highlights the  scarcity  of bulge-less  galaxies
  (i.e. low $n$) at high masses.}
\label{fig:fs}
\end{center}
\end{figure*}

\begin{figure*}
\centerline
{\psfig{figure=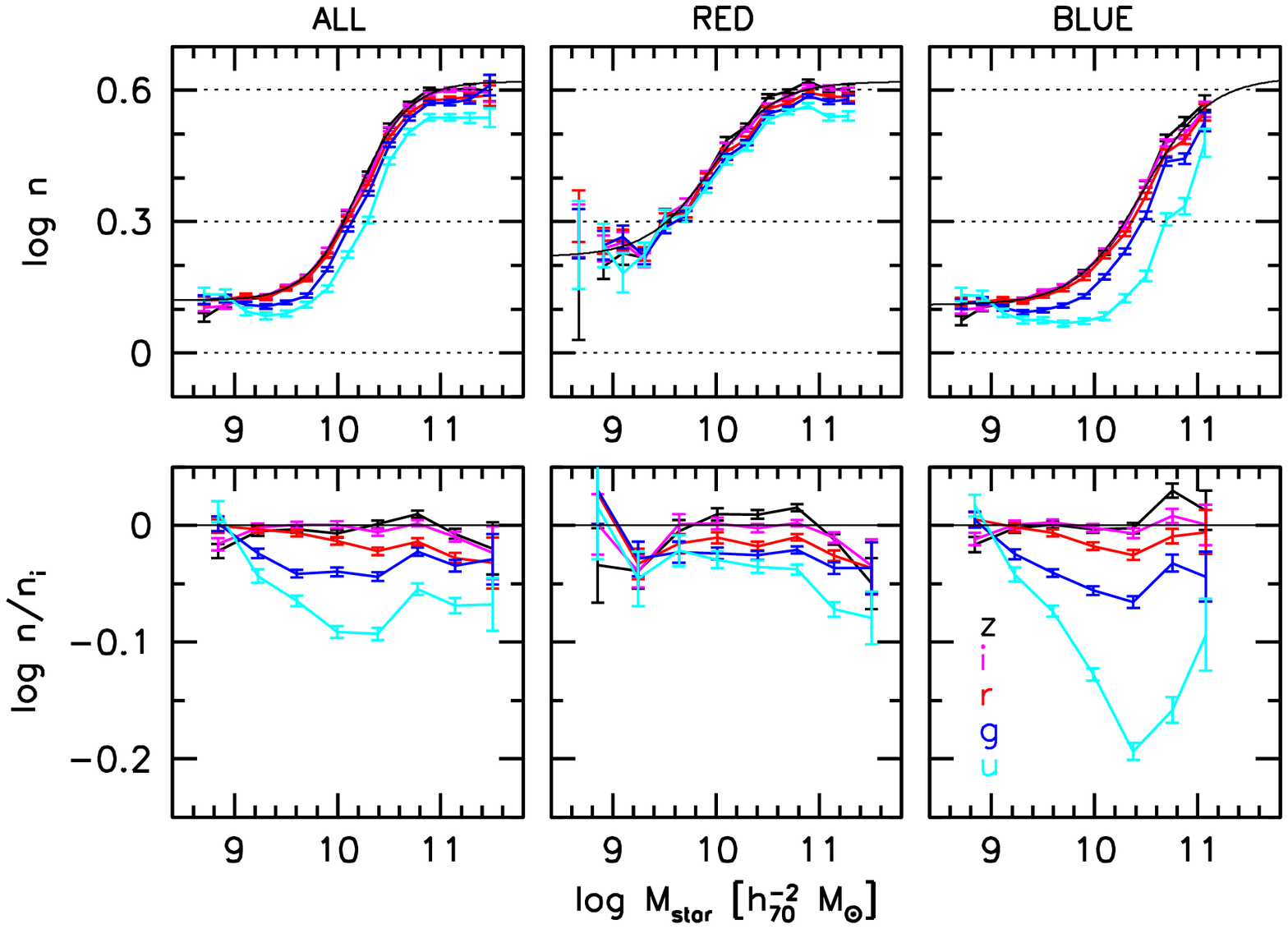,width=0.80\textwidth}}
\caption[   ]{\footnotesize   Dependence    of   S\'ersic   index   on
  pass-band. The upper panels show the median S\'ersic index - stellar
  mass relation for  S\'ersic index measured in each  of the five SDSS
  pass-bands: $u,g,r,i,$ and $z$.  The lower panels show the deviation
  of  the  S\'ersic index  measured  in  the  $u,g,r,$ and  $z$  bands
  relative  to  that measured  in  the  $i$  band. The  dependence  of
  S\'ersic index on wavelength is strongest for blue galaxies, and peaks at the transition mass of $\Mstar \simeq 2.5 \times 10^{10}\Msun$. }
\label{fig:cn5}
\end{figure*}

\section{How Common are Bulge-Less Exponential Disks?}
We have shown  that it is possible to  produce quasi-exponential disks
with \LCDM initial conditions.  However, these galaxies are not
typical (in the sense they do not occur for median values of the input
parameters)  for  any  halo  mass  between  $10^{11}$  and  $10^{13}$.
However, in our  models, low mass haloes produce  disks with closer to
exponential  stellar  profiles.    Thus,  an  important  observational
question  is to  determine what  fraction of  galaxies,  and late-type
galaxies in particular, have bulge-less exponential disks and low
bulge fractions, as a function of stellar mass.

It  is generally  thought  that bulge-less  disk  galaxies are  fairly
common, and as such, this  presents a challenge to hierarchical galaxy
formation  models.   This is  especially  the  case  if a  significant
fraction  of galaxy  bulges were  formed via  secular  processes (e.g.
Kormendy \&  Fisher 2005).  Such  a conclusion is supported  by recent
analyses  that find  bulge-less disk  galaxy fractions  of 15\%  for a
sample of edge-on disk galaxies  (Kautsch \etal 2006), and $\sim 20\%$
for sample  of visually classified  quasi-bulge-less galaxies (Barazza
\etal  2008).   However,  these  analyses  do not  consider  any  mass
dependence to the fraction of bulge-less disk galaxies, which may lead
one to over predict the frequency of massive bulge-less galaxies.

In an attempt to get at least a preliminary (if not a definitive)
answer to the question of how common are exponential bulge-less disks
(as a function of stellar mass) we make use of S\'ersic indices from
the publicly available low redshift New York University Value Added
Catalog (NYU-VAGC, Blanton \etal 2005).  The low-$z$ NYU-VAGC is based
on the second data release of the SDSS (Abazajian \etal 2004), and
consists of 28089 galaxies at distances of 10-200 Mpc ($0.0033 < z <
0.05$), which have been determined by correcting for peculiar
velocities.

The S\'ersic profile is given by
\begin{equation}
  \Sigma(R) = \Sigma_0 \exp[ -(R/R_{\rm s})^{1/n}],
\end{equation} 
where $\Sigma_0$  is the central  surface density, $R_{\rm s}$  is the
scale radius, and  $n$ is the S\'ersic index.   For $n=4$ this profile
corresponds  to a de  Vaucouleurs profile  (de Vaucouleurs  1959), for
$n=1$ to  an exponential profile, and  for $n=0.5$ to  a gaussian. The
NYU-VAGC provides S\'ersic  indices measured in each of  the five SDSS
pass-bands: $ugriz$.   The S\'ersic  fits are performed  with circular
apertures, and thus  are likely to given erroneous  results for highly
inclined  disks.   Furthermore,  the  surface brightness  profiles  of
highly inclined disks are often strongly affected by extinction, so we
wish to remove  inclined disks remove from our sample.   To do this we
use the  axis ratios  measured from exponential  fits (taken  from the
main SDSS catalog), and remove  galaxies with axis ratios smaller than
$b/a=0.5$   (i.e.   corresponding   to  an   inclination   of  $\simeq
60^{\circ}$).  This removes about half the galaxies in the sample.  De
Vaucouleurs fit  axis ratios are  also available, but these  often give
axis ratios more  representative of the bulge or  bar, rather than the
disk, so they are not considered useful for our purposes.

For both S\'ersic  index and axis ratio we adopt  the $i$-band fits as
this provides a trade off  between signal to noise and wavelength.  We
calculate stellar  masses using the  relation between $g-r$  color and
stellar mass-to-light ratio from Bell  \etal (2003), with an offset of
-0.1 dex corresponding to a Chabrier (2003) IMF:
\begin{equation}
  \log \Mstar/L_{\rm r} = -0.406 +1.097 (g-r)
\end{equation}

The upper panels in  Fig.~\ref{fig:cn} show the relation between color
and stellar  mass.  We  separate galaxies into  red and  blue galaxies
using $u-r$  colors following Baldry  \etal (2004), who  determine the
optimal color separation between red and blue galaxies to be
\begin{equation}
(u-r)_{\rm div} = 2.06 -0.244 \,\tanh \left( \frac{M_r+20.07}{1.09}\right),
\end{equation}
where  $M_{\rm  r}$ is  the  $r$-band  magnitude  of the  galaxy.   In
addition,  to keep  the red-sequence  and blue-cloud  ``pure''  in the
$g-r$ vs  $\Mstar$ plane, we require  that red galaxies  are not bluer
than  0.1 dex from  the red-sequence,  and that  blue galaxies  are at
least 0.05 dex bluer  than the red-sequence.  The color-mass relations
for red  and blue  galaxies are  shown in the  upper middle  and upper
right panels respectively.

The  lower  panels in  Fig.~\ref{fig:cn}  show  the relations  between
S\'ersic index (measured  in the $i$-band) and stellar  mass.  For all
galaxies  (left panel) the  S\'ersic index  also shows  a bi-modality,
with a cloud of points at  high masses around $n=4$, and at low masses
around $n=1.3$. There is a  rapid transition in S\'ersic index between
a   stellar  mass   of  $\Mstar   =  1\times   10^{10}$   to  $5\times
10^{10}\Msun$.  It is tempting  to associate this bi-modality with the
color  bi-modality.   However, if  we  look  at  early and  late-types
separately,  we  see the  same  trends:  low  mass galaxies  have  low
S\'ersic index  and high  mass galaxies have  high S\'ersic  index. We
return  to  the  correlation  between  color  and  S\'ersic  index  in
\S~\ref{sec:colser} below.

In order to  calculate residuals, and to ease  comparisons with models
we fit the relations in  Fig.~\ref{fig:cn} with the following.  We fit
the color-mass relation of the red galaxies with a single power-law
\begin{equation}
\label{eq:colred}
(g-r) = (g-r)_0 + \log \left(\frac{\Mstar}{M_0}\right)^\alpha.
\end{equation}
We fit the color-mass relation of the blue galaxies with a double power-law
\begin{equation}
\label{eq:colblue}
(g-r) = (g-r)_0 + \log \left(\frac{\Mstar}{M_0}\right)^\alpha + \log \left(\frac{M_0+\Mstar}{2M_0}\right)^{\beta-\alpha}.
\end{equation}
We fit the S\'ersic index - stellar mass relations with the following function:
\begin{equation}
\label{eq:sersic}
\log n = \log n_2 + \frac{(n_1 - n_2)}{1+10^{\gamma \log (\Mstar/M_0)}}.
\end{equation}
Here $n_1$  is the asymptotic value of  $n$ at low mass,  $n_2$ is the
asymptotic value  of $n$ at high  mass, $M_0$ is  the transition mass,
and $\gamma$  controls the  sharpness of the transition.   The best
fit parameters are given in table \ref{tab:params}.

\begin{table}
 \centering
 \begin{minipage}{70mm}
   \caption{Best   fit   parameters    for   median   color-mass   and
     S\'ersic-mass   relations   from   Fig.~\ref{fig:cn}  using   the
     functions in Eqs.~(\ref{eq:colred}-\ref{eq:sersic}).}
  \begin{tabular}{lcccc}
\hline 
\hline  
Sample & $\log M_0$ & $(g-r)_0$ & $\alpha$ \\
RED    & 10.00 & 0.718 & 0.067 & \\
\hline  
Sample & $\log M_0$ & $(g-r)_0$ & $\alpha$ & $\beta$ \\
BLUE   & $\phantom{1}9.10$ & 0.375 & 0.00 & 0.182 \\
\hline  
Sample & $\log M_0$ & $\log n_1$ & $\log n_2$ & $\gamma$ \\
ALL & 10.16 & 0.12 & 0.62 & 1.95\\
RED & $\phantom{1}9.90$ & 0.21 & 0.62 & 1.70 \\ 
BLUE& 10.41 & 0.12 & 0.61 & 1.70 \\ 
\hline 
\hline
\label{tab:params}
\end{tabular}
\end{minipage}
\end{table}

From Fig.\ref{fig:cn}  it is apparent  that massive galaxies  with low
S\'ersic indices are  rare, both for red and  blue sub-samples.  To be
more quantitative Fig.~\ref{fig:fs} shows  the fraction of galaxies at
a given  stellar mass  with S\'ersic index  less than 1.0  (cyan), 1.5
(blue), 2.0 (green), 2.5 (red),  and 4.0 (magenta).  Also shown is the
fraction of all galaxies that  are red (black dotted).  The red-galaxy
fraction steadily increases from low  to high stellar mass.  Note that
our red  galaxy fraction  is biased  high due to  our axis  ratio cut,
which preferentially removes disk (and hence typically blue) galaxies.

At the transition mass of $2.5\times 10^{10}\Msun$ the median S\'ersic
index of the late types is  $n\simeq 2$.  We adopt a S\'ersic index of
$n=1.5$ as an upper-limit on the S\'ersic index of a bulge-less galaxy
(which  was  also the  definition  used by  Bell  2008).  Thus at  the
transition  mass,  less  than  $\simeq  15\%$  of  blue  galaxies  are
bulge-less.   At higher  masses the  bulge-less galaxy  fraction drops
rapidly.  To show  just how rare massive bulge-less  galaxies are, the
lower panels show the fractions on a logarithmic scale.

Using  the TF  relations from  Pizagno \etal  (2005) and  Dutton \etal
(2007), galaxies with rotation velocities  of 220 $\kms$ (e.g. such as
the  Milky-Way)   have  a  median  stellar  mass   of  $\Mstar  \simeq
10^{11}\Msun$. Note that  the actual stellar mass of  the Milky-Way is
thought to  be $\simeq 5\times 10^{10}\Msun$ (Widrow,  Pym \& Dubinski
2008),  which  means   that  the  Milky-Way  does  not   fall  on  the
TF  relation  (Hammer \etal  2007).  These  galaxies have  a
median S\'ersic index  of $n\simeq 3.5$, and less  than 0.1\% of these
galaxies  are  bulge-less (i.e.  have  $n<1.5$).   This provides  some
relief for theorists who have  been trying unsuccessfully to produce a
bulge-less Milky-Way  mass galaxy  (e.g.  Abadi \etal  2003; Robertson
\etal 2004; Governato \etal 2004; 2007).  A more serious challenge for
theorists is to produce galaxies like M33, as the majority of galaxies
at these  masses do  not have significant  bulges.  For a  galaxy like
M33, with a rotation velocity of 120$\kms$ and a stellar mass of $1\times
10^{10}\Msun$,  the median $n\simeq  1.6$ and  45\% of  blue galaxies
have $n<1.5$.   This means that quasi-bulgeless galaxies  are the norm
in low mass galaxies, which means they must form in dark matter haloes
with   typical  mass   accretion  histories.   Thus   if  cosmological
simulations fail to produce galaxies  like M33, this signals a failure
of the model, and not an unfortunate choice of mass accretion history.
By  contrast,  since massive  disk  dominated  galaxies,  such as  the
Milky-Way (which has a bulge  fraction of $\simeq 0.2$ (Widrow, Pym \&
Dubinski 2008)  are rare, they may  have formed in haloes  with a very
unusual mass accretion histories.  This makes simulating the Milky-Way
a  harder problem,  because  it would  take  a sample  of hundreds  of
Milky-Way  mass haloes  before one  would expect  to produce  a galaxy
resembling the Milky-Way.

\begin{figure*}
\centerline
{\psfig{figure=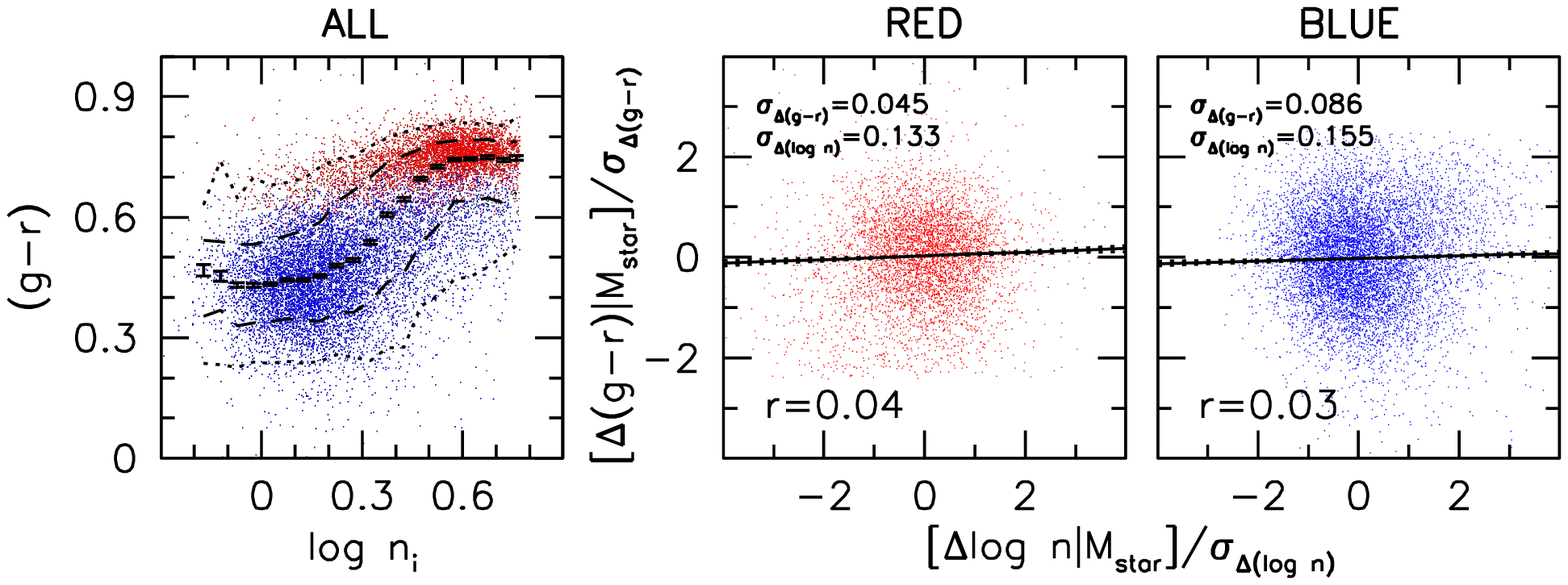,width=0.80\textwidth}}
\centerline
{\psfig{figure=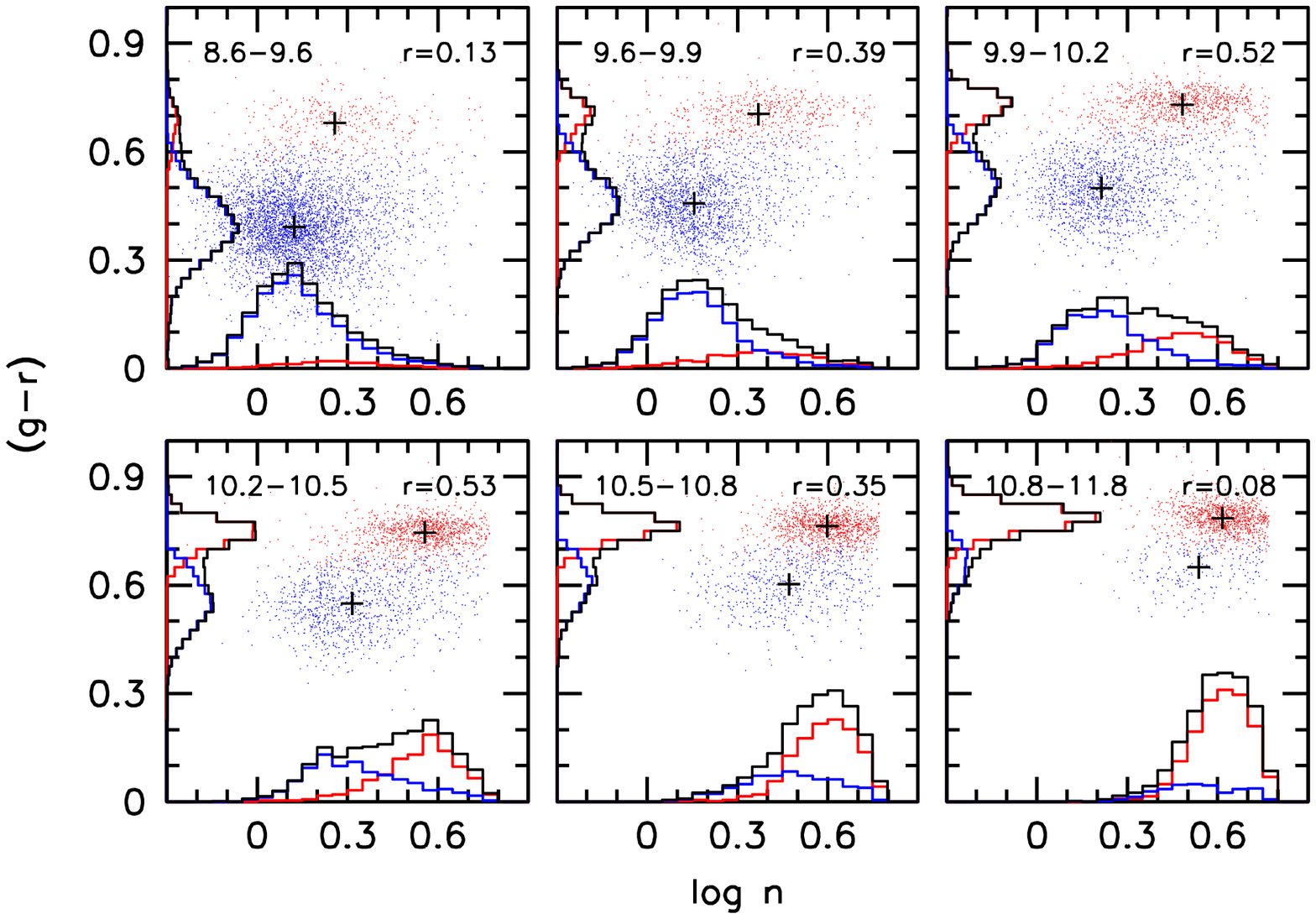,width=0.80\textwidth}}
\caption[  ] {\footnotesize  Correlations between  color  and S\'ersic
  index.  The upper  left panel shows color vs  S\'erisc index for all
  galaxies in our  sample, color coded by whether they  are in the red
  sequence  or blue  cloud.  This  shows a  clear  correlation between
  color and S\'ersic index.  The upper right panels show the residuals
  of the color-mass relation versus the residuals of the S\'ersic mass
  relation, both for  red and blue galaxies. The  solid lines show the
  best fit of  $\Delta$ color vs $\Delta$ S\'ersic  index, with dashed
  lines  indicating   the  error   on  the  slope.    The  correlation
  coefficient,  $r$, for  each fit  is  also shown.   This shows  that
  within  the blue cloud  or red  sequence, the  scatter in  color and
  S\'ersic  index, at  a given  stellar mass,  are  uncorrelated.  The
  lower panels show the color-S\'ersic relations for 6 mass bins.  The
  range in  $\log \Mstar/[h_{70}^{-2}\Msun]$ for each bin  is given in
  the  top left  corner of  each panel.   The  correlation coefficient
  between color and S\'ersic index is given in the top right corner of
  each panel. The median color and S\'ersic index for each mass bin is
  indicated by  a cross.  This shows  that whether a galaxy  is red or
  blue  does  depend  on  its  S\'ersic  index,  especially  near  the
  transition mass of $\Mstar\simeq 2.5\times 10^{10}\Msun$.}
\label{fig:scatter}
\end{figure*}

\begin{figure*}
\centerline{\psfig{figure=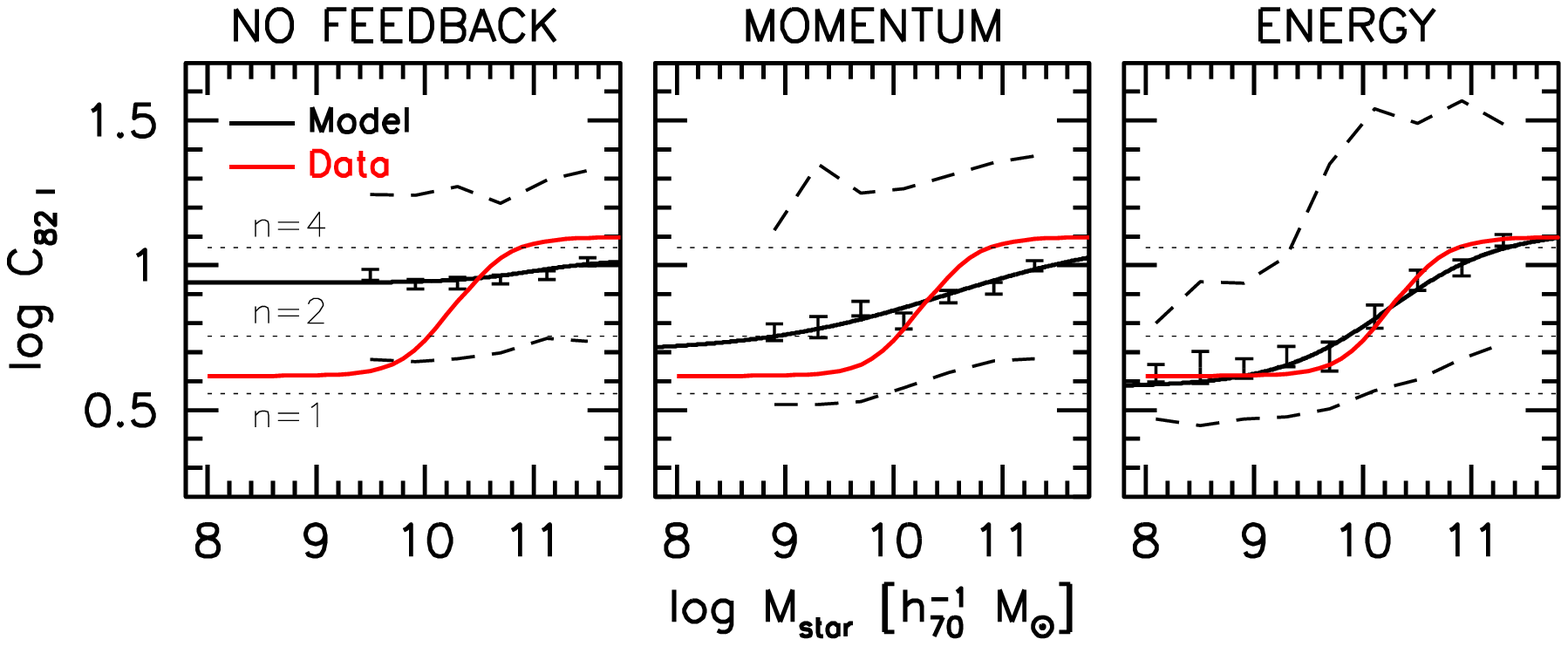,width=0.80\textwidth}}
\caption[  ]{\footnotesize Concentration  of the  I-band  light versus
  stellar mass  for three feedback  models.  The black lines  show the
  16th, and 84th percentiles (dashed),  and a fit to the median values
  (solid). The red line shows the median relation for the observations
  (from  Fig.~\ref{fig:cn}  after converting  from  S\'ersic index  to
  concentration.    The  dotted   lines  show,   for   reference,  the
  concentration corresponding to a S\'ersic index of $n=1,2,$ and $4$.
  The energy  driven feedback model  produces disk galaxies  that have
  S\'ersic indices that are in good agreement with the observations.}
\label{fig:cm4}
\end{figure*}

\subsection{The Wavelength Dependence of the S\'ersic Index}
In Fig.~\ref{fig:cn5} we show the median S\'ersic index - stellar mass
relation  for  the S\'ersic  index  measured in  each  of  the 5  SDSS
pass-bands $ugriz$.   The general trend is for  lower S\'ersic indices
in bluer pass bands. This is  true for both red and blue galaxies, but
the  effects are  strongest  for  the blue  galaxies.   This could  be
attributed  to   the  greater  sensitivity  of   bluer  pass-bands  to
extinction  and/or  variations in  the  star  formation  history as  a
function   of   galacto   centric   radius.  The   lower   panels   of
Fig.~\ref{fig:cn5} show  the differences between the  S\'ersic index as
measured  in  the $u,g,r,$  and  $z$-bands  to  that measured  in  the
$i$-band.  The largest differences occur around the transition mass of
$2.5\times 10^{10}\Msun$.

These results have implications for studies of galaxy structure at
high redshift.  Currently large surveys with HST imaging, such as the
Galaxy Evolution and Morphology Survey (GEMS) (Rix \etal 2004), the
All-wavelength Extended Groth strop International Survey (AEGIS)
(Davis \etal 2007), and the cosmological evolution survey (COSMOS)
(Scoville \etal 2007; Koekemoer \etal 2007) are limited to the F814W
or the F850LP filters (hereafter referred to as the $I814$ and $z850$
bands).  At redshift $z=1$ the $I814$ filter shifts into the $U$-band,
while the $z850$ filter corresponds to B-band.  In an attempt to make
a fair comparison between galaxies at different redshifts, authors
often compare structural properties in the rest frame $B$ or $V$ bands
at low and high redshifts.  However, given that these bands do not
trace stellar mass even at low redshift, and that star formation rates
at $z \simeq 1$ were an order of magnitude higher than they are today
(e.g.  Noeske \etal 2007), it is not clear that such a comparison
gives an unbiased view of the evolution of galaxy structure in stellar
mass.  To determine the structural properties of galaxies at $z\gta 1$
free of the biases of young stellar populations and dust will require
high resolution NIR imaging, preferably in more than one band to a
provide rest frame red pass band and an optical color to correct for
stellar population and extinction gradients within galaxies.

\subsection{Is S\'ersic Index Correlated with Color?}
\label{sec:colser}
From  Fig.~\ref{fig:cn} we  see  that higher  mass  galaxies are  both
redder and more concentrated  than lower mass galaxies.  Thus globally
there is a correlation between S\'ersic index and color: galaxies with
lower S\'ersic  index are bluer;  galaxies with higher  S\'ersic index
are  redder.  This correlation  is shown  in the  upper left  panel of
Fig.~\ref{fig:scatter}. However,  this does not  mean that there  is a
causal  relation between S\'ersic  index and  color.  The  upper right
panels of Fig.~\ref{fig:scatter} show  the residuals of the color-mass
relation vs.   the residuals of the S\'ersic-mass  relation.  For both
red  and  blue  galaxies  the  correlation is  consistent  with  zero,
i.e. within the blue cloud or red-sequence, the scatter in color, at a
given  stellar mass,  is  uncorrelated with  the  scatter in  S\'ersic
index.   How  do  we  reconcile  this  fact  with  the  strong  global
correlation between S\'ersic index and color?

The  middle  and  lower   panels  of  Fig.\ref{fig:scatter}  show  the
color-S\'ersic  relation for  6  mass bins  from $\Mstar=10^{8.6}$  to
$10^{11.8}\Msun$. This  shows that at  a given stellar mass  whether a
galaxy is  blue or red does  depend on its  S\'ersic index, especially
near the  transition mass of $2.5\times  10^{10}\Msun$.  Galaxies with
high S\'ersic indices are more likely to be on the re-sequence than in
the  blue cloud.   This result  is  consistent with  Bell (2008),  who
showed that having a bulge is a requirement for a central galaxy being
on the  red-sequence, which  is a central  prediction of  AGN feedback
scenarios for quenching star formation.

\begin{figure*}
\centerline{\psfig{figure=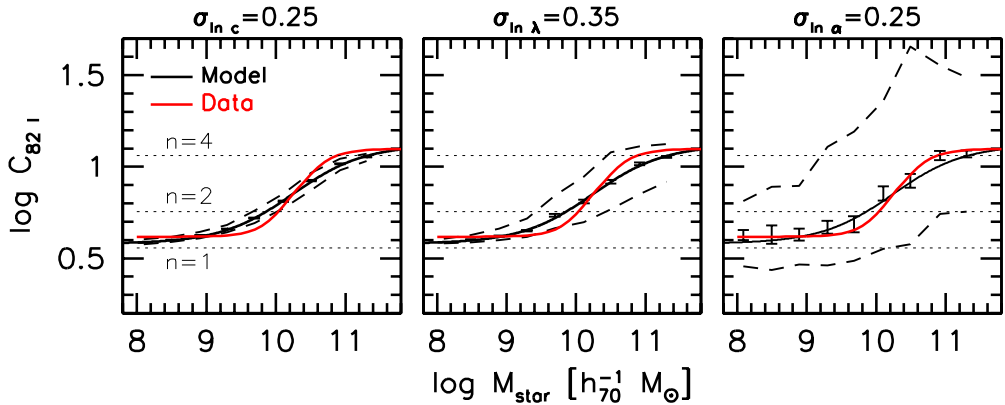,width=0.80\textwidth}}
\caption[ ]{\footnotesize  Concentration-mass relation for  the energy
  feedback model  for three different sources of  scatter: $c$ (left);
  $\lambda$  (middle)  and  $\alpha$  (right).   The  scatter  in  the
  concentration mass relation is dominated by scatter in $\alpha$, the
  angular   momentum   shape   parameter.   The  lines   are   as   in
  Fig~\ref{fig:cm4}.}
\label{fig:cm5}
\end{figure*}

\section{Comparison between Models and Observations}
Having  established observationally  that S\'ersic  index is  a strong
function of  stellar mass, we  now investigate whether our  models can
reproduce this trend. We consider three models with different feedback
prescriptions:  1) no feedback;  2) momentum  driven feedback;  and 3)
energy driven feedback.   The parameters of these models  are given in
Table~\ref{tab:models}.   For  each  model  we  generate  Monte  Carlo
samples  with halo  masses ranging  from  $10^{10} <  \Mvir <  10^{13}
\hMsun$,  corresponding to  virial  velocities ranging  from $31  \lta
\Vvir \lta 310 \kms$.  We  also include log-normal scatter in the halo
concentration, $c$,  halo spin parameter, $\lambda$,  and halo angular
momentum shape  parameter, $\alpha$.   The scaling relations  of these
models are discussed in more detail in DB08.

\begin{figure*}
\centerline{\psfig{figure=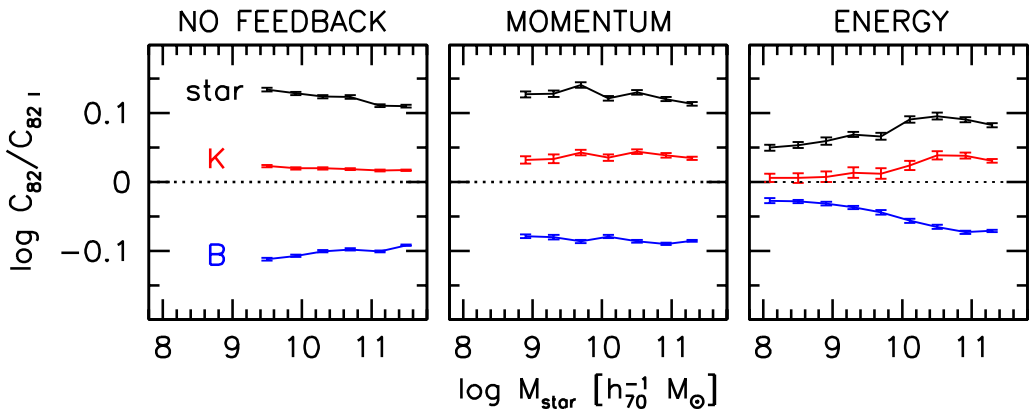,width=0.80\textwidth}}
\caption[ ]{\footnotesize Difference between concentration measured in
  I-band to that measured in  stellar mass (black), K-band light (red)
  and  B-band  light  (blue),  vs  stellar mass.   The  energy  driven
  feedback model produces disk galaxies that have color gradients that
  are in good agreement with the observations.}
\label{fig:cm4b}
\end{figure*}

\subsection{Concentration Mass Relation}

The upper panels of  Fig.~\ref{fig:cm4} shows the concentration of the
I-band  light  as   a  function  of  stellar  mass   for  these  three
models. Recall that we define the concentration parameter as the ratio
between  the  radii  enclosing  80  and  20\%  of  the  I-band  light:
$C_{82I}=R_{80I}/R_{20I}$.

The  model  without feedback  results  in  a  median concentration  of
$\simeq 9$,  which corresponds  to a S\'ersic  index of  $\simeq 3.5$.
The  high  concentrations  are  a  result  of  the  high  galaxy  mass
fractions,  which cause  the baryons  to dominate  the  inner circular
velocity profile.   The model with  momentum driven wind  also results
high   concentrations    at   high   stellar    masses   $\Mstar\simeq
10^{11}\Msun$, but also shows a trend of decreasing concentration with
decreasing   stellar   mass,  in   qualitative   agreement  with   the
observations (solid  red line).  However, the observed  relation has a
sharper transition, as well as both lower concentrations at low masses
and  higher concentrations  at high  masses.  The  energy  driven wind
model results in  a much better agreement with  the observed relation.
It has  the correct asymptotic  values for the concentrations  at both
low and  high masses.  In particular this  model successfully produces
galaxies with close to  exponential surface brightness profiles at low
stellar masses. We emphasize that none of these models have been tuned
to reproduce  the concentration mass data.   The feedback efficiencies
of  these models were  chosen so  that the  models reproduce  the zero
points  of  the  TF  and size-stellar  mass  relations  (see
DB08).

\begin{table}
 \centering
 \begin{minipage}{140mm}
  \caption{Model Parameters}
  \begin{tabular}{lccccc}
\hline  Name &  $\epsilon_{\rm EFB}$ & $\epsilon_{\rm MFB}$ & $\bar{\lambda}$& $\sigma_{\ln \lambda}$ & AC \\
I: No Feedback: & 0.0 & 0.0 & 0.035 & 0.35 & N \\ 
II: Momentum:& 0.0 & 1.0 & 0.035 & 0.35 & N \\ 
III: Energy: & 0.25 & 0.0 & 0.025 & 0.35 & N \\ 
\hline 
\label{tab:models}
\end{tabular}
\end{minipage}
\end{table}

\subsection{Scatter in Concentrations}
In  all   three  models  there   is  a  significant  scatter   in  the
concentration index at a given  stellar mass.  To determine where this
scatter    comes   from    we   show    in    Fig.~\ref{fig:cm5}   the
concentration-mass relation for the energy driven feedback model where
there  is only  one source  of scatter:  $c$, $\lambda$,  or $\alpha$.
This shows  that the scatter in the  I-band concentration-stellar mass
relation is  dominated by the  angular momentum shape  parameter, with
the  halo   spin  parameter   contributing  some  scatter,   and  halo
concentration resulting in a  negligible amount of scatter.  Note also
that the scatter in concentrations  depends on stellar mass, with less
scatter  at lower masses.   This shows  that feedback  is able  to, at
least partially, erase the initial conditions.

The  scatter in  the  concentration  mass relation  in  the models  is
significantly  larger,  by at  least  a factor  of  two,  than in  the
observations.   In  Fig.~\ref{fig:sv-BS}  we  saw that  the  SS05  AMD
results  in larger  scatter in  disk concentration  that the  B01 AMD.
Thus an important question for theorists is to determine in detail the
range of AMD's of dark matter and gas in cosmological simulations.

\subsection{Differences  Between Concentrations  in  Stellar Mass  and
  Light}
\label{sec:colser2}
Fig.\ref{fig:cm4b} shows  the differences,  with respect to  the I-band
light,  of the  concentration  indices measured  in  the stellar  mass
(black  lines),  K-band  light  (red  lines) and  B-band  light  (blue
lines). As might  be expected, in all three  models the concentrations
are highest in stellar mass, and lowest in B-band light.  As discussed
in \S~\ref{sec:stelpop}  the color gradients  are a reflection  of the
inside-out  formation   of  stellar  disks.   However,   what  may  be
surprising is that the K-band concentrations are not equal to those of
the  stellar  mass. Furthermore,  the  difference  between I-band  and
stellar  mass is about  the same  as that  between B-band  and I-band.
This suggests  that an accurate determination of  the concentration of
the stellar  mass in disk  galaxies requires the color  gradients (and
hence stellar mass-to-light gradients) to be taken into account.

In  the observations, low  mass galaxies,  $\Mstar\simeq 10^{9}\Msun$,
have  small differences  between  S\'ersic indices  in different  pass
bands.   The differences  between S\'ersic  indices in  different pass
bands increase  with stellar  mass, reaching a  maximum at  around the
transition  mass of  $2.5\times 10^{10}\Msun$.   Qualitatively similar
trends of  light concentration  with stellar mass  are found  with our
energy feedback model.  The  momentum feedback and no feedback models,
however,  have almost  no mass  dependence to  the  concentrations, in
clear conflict with the observations.

\begin{figure*}
\centerline
{\psfig{figure=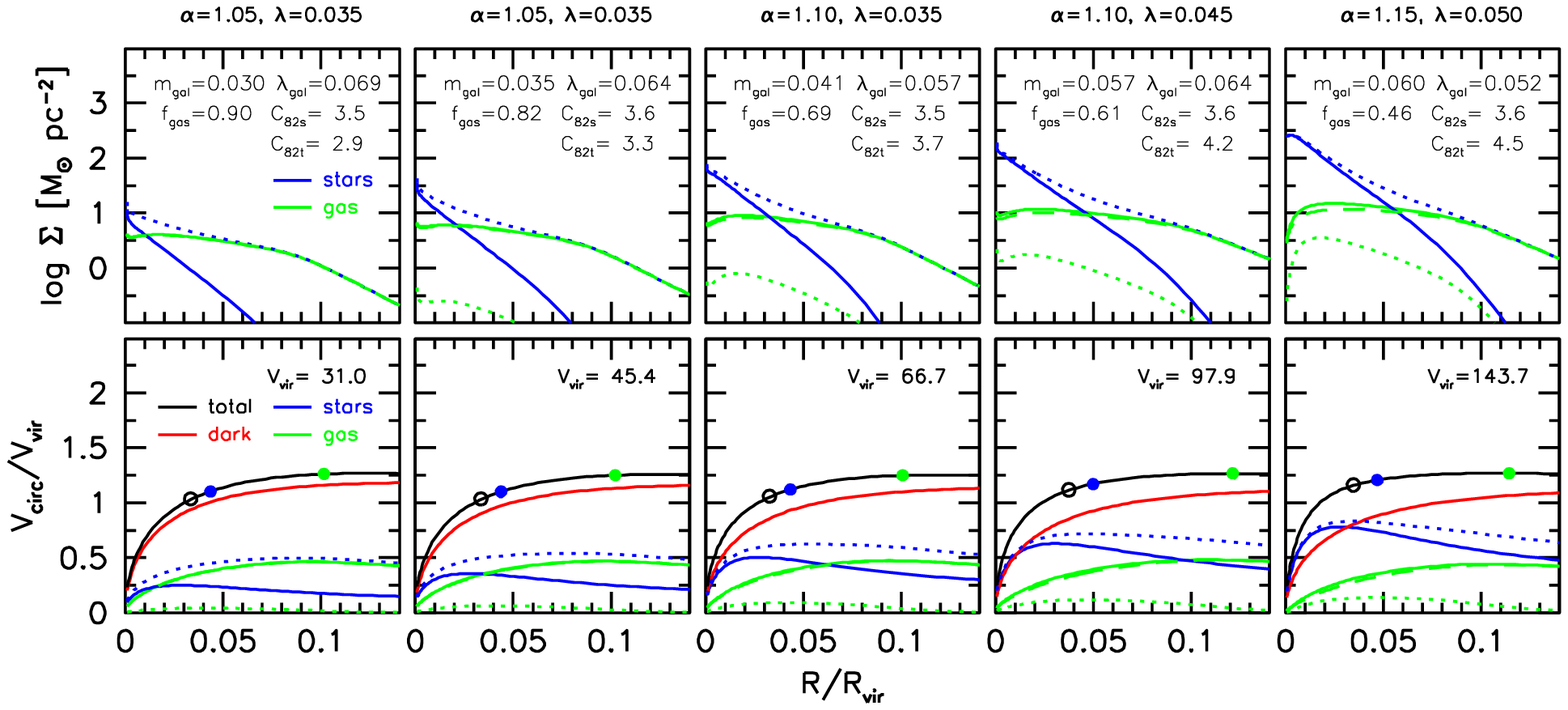,width=0.80\textwidth}}
\centerline
{\psfig{figure=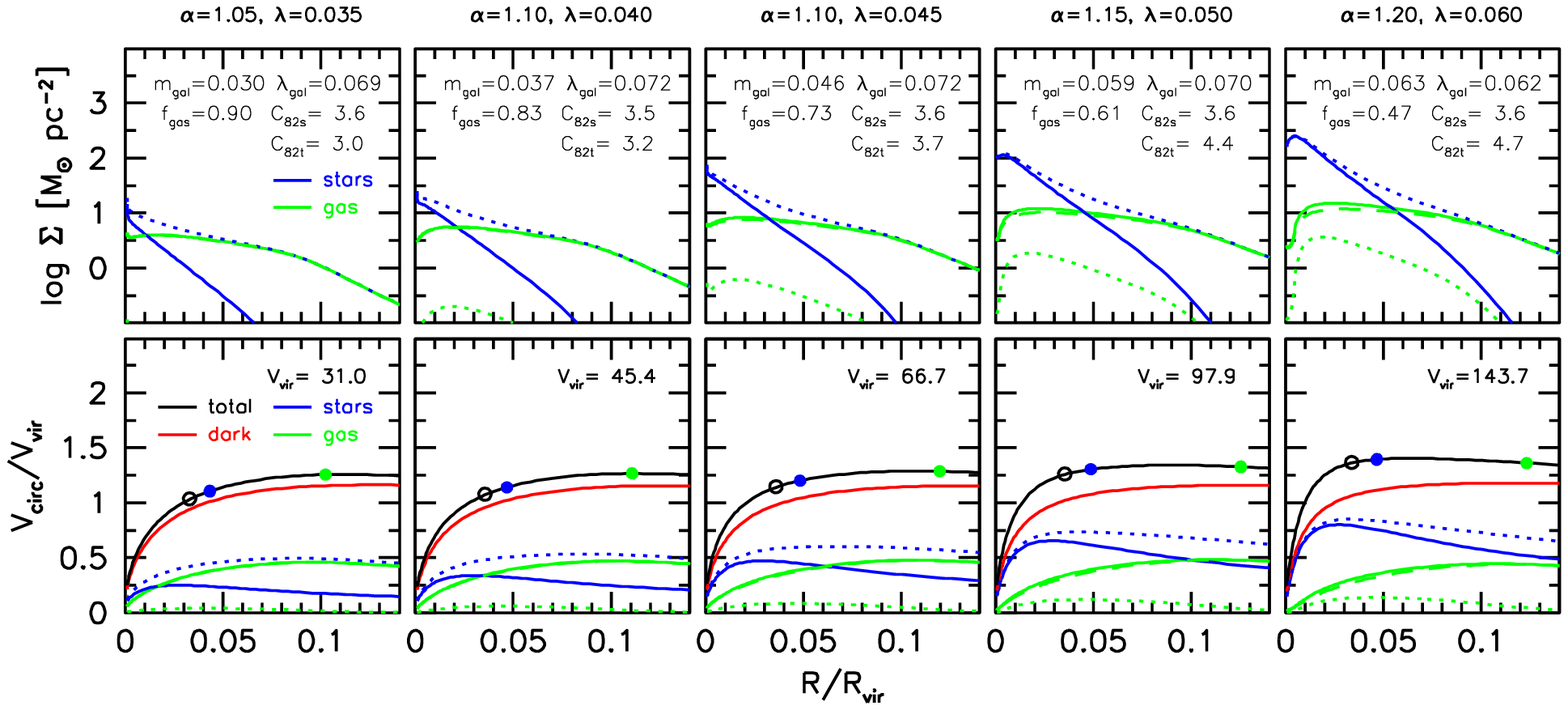,width=0.80\textwidth}}
\caption[]{\footnotesize Examples of galaxies with exponential stellar
  density profiles (solid blue lines) in our energy driven feedback
  ($\epsilon_{\rm EFB}=0.25$) model.  Virial mass increases left to
  right from $\Mvir=10^{10}\hMsun$ to $\Mvir=10^{12}\hMsun$.  All
  models follow the concentration mass relation. The halo angular
  momentum parameters, $\alpha$ and $\lambda$, for each model are
  specific at the top of the surface density panels. These values are
  typically within 1$\sigma$ of the values predicted from cosmological
  simulations.  The models produce exponential stellar disks with
  (lower panels) and without (upper panels) adiabatic contraction. The
  line types and colors are the same as in Fig.\ref{fig:sv-BS}.}
\label{fig:sv-exp}
\end{figure*}

\subsection{Examples of Model Exponential Stellar Disks}
\label{sec:exp}
From the light concentration - stellar mass plots of
Figs.\ref{fig:cm4} \& \ref{fig:cm5} it is apparent that galaxies with
concentrations consistent with pure exponential surface brightness
profiles are produced by our models with feedback, typically within
the $1\sigma$ variations of the halo angular momentum parameters. In
Fig.\ref{fig:sv-exp} we show examples of such cases. These models have
exponential stellar disks over the radial range of $\simeq 0$ to
$\simeq 5$ stellar disk scale lengths. Virial mass increases left to
right from $\Mvir=10^{10}\hMsun$ to $\Mvir=10^{12}\hMsun$. The halo
angular momentum parameters ($\alpha,\lambda$) of these models are
specified above the surface density plots. For comparison, the median
and 1$\sigma$ upper scatter in these values in cosmological
simulations are (0.90,1.15) for $\alpha$ (Sharma \& Steinmetz 2005)
and (0.035, 0.60) for $\lambda$ (Maccio \etal 2007). The upper panels
are galaxies taken from Model III, while the lower panels are taken
from an equivalent model but with adiabatic contraction of the
halo. The models with adiabatic contraction also produce pure
exponential stellar disks, albeit with slightly higher values of
($\alpha$ and $\lambda$). This demonstrates that the ability of our
outflow models to produce pure exponential stellar disks is not
dependent on the effect of adiabatic contraction.

\begin{figure*}
\centerline
{\psfig{figure=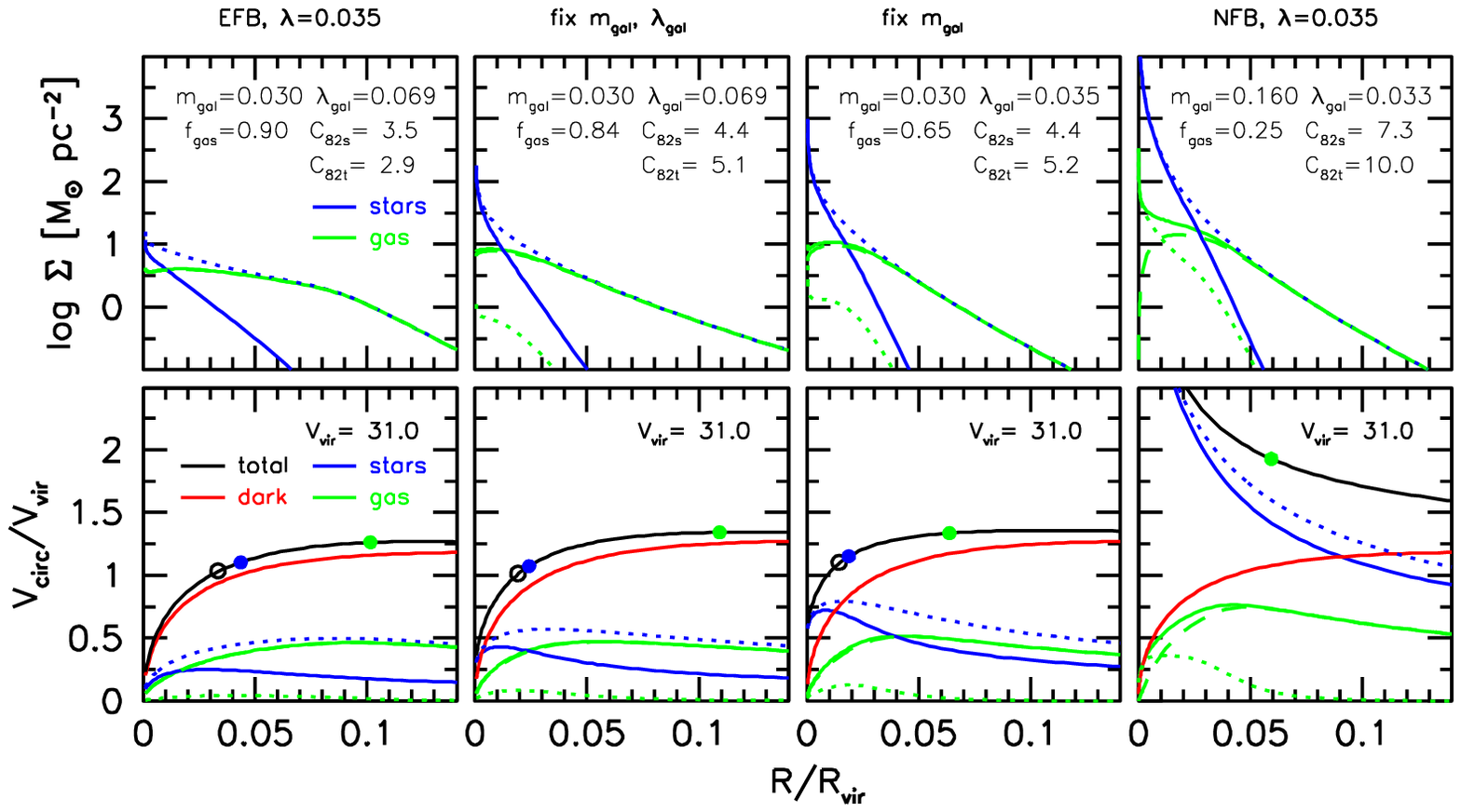,width=0.67\textwidth}}
\centerline
{\psfig{figure=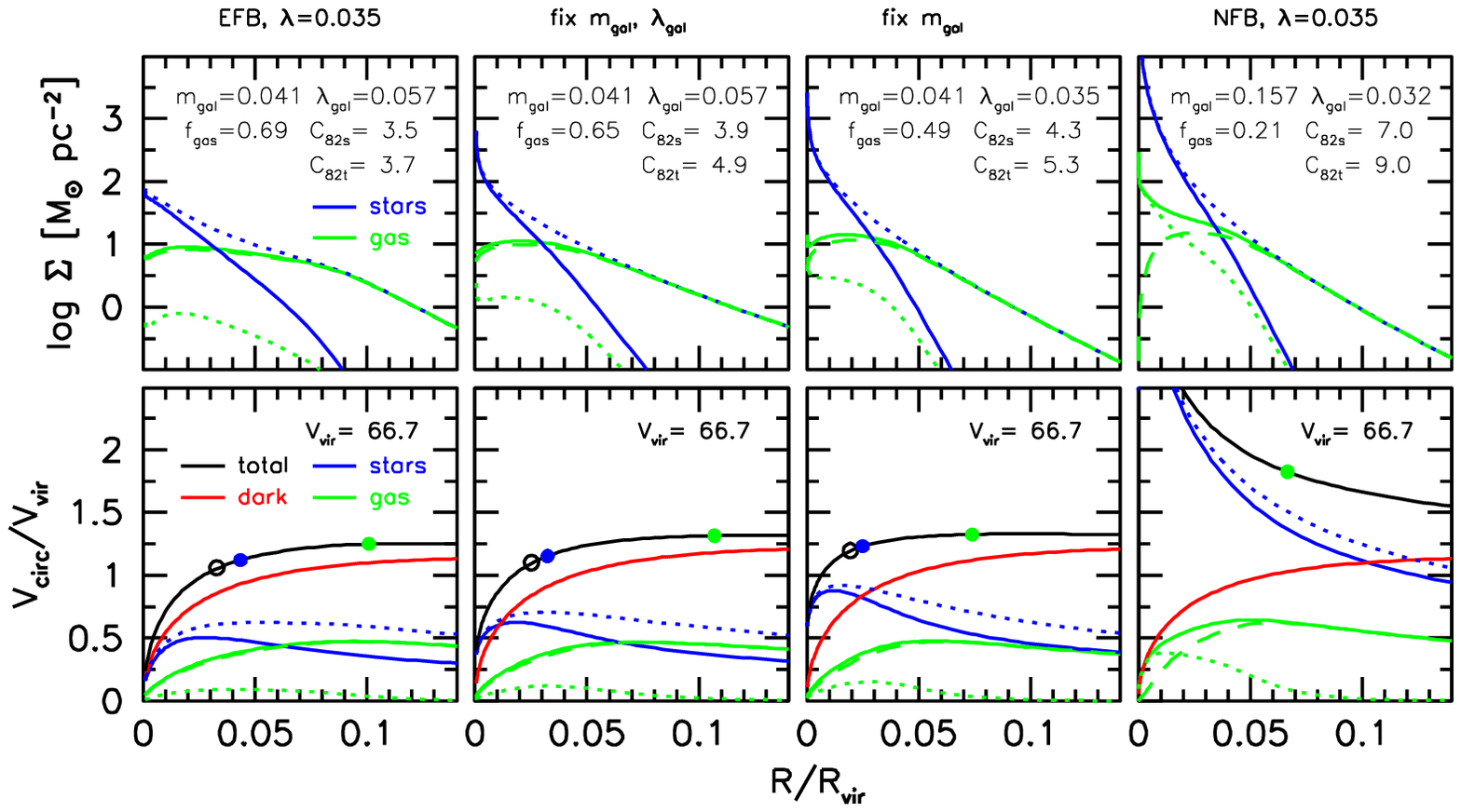,width=0.67\textwidth}}
\centerline
{\psfig{figure=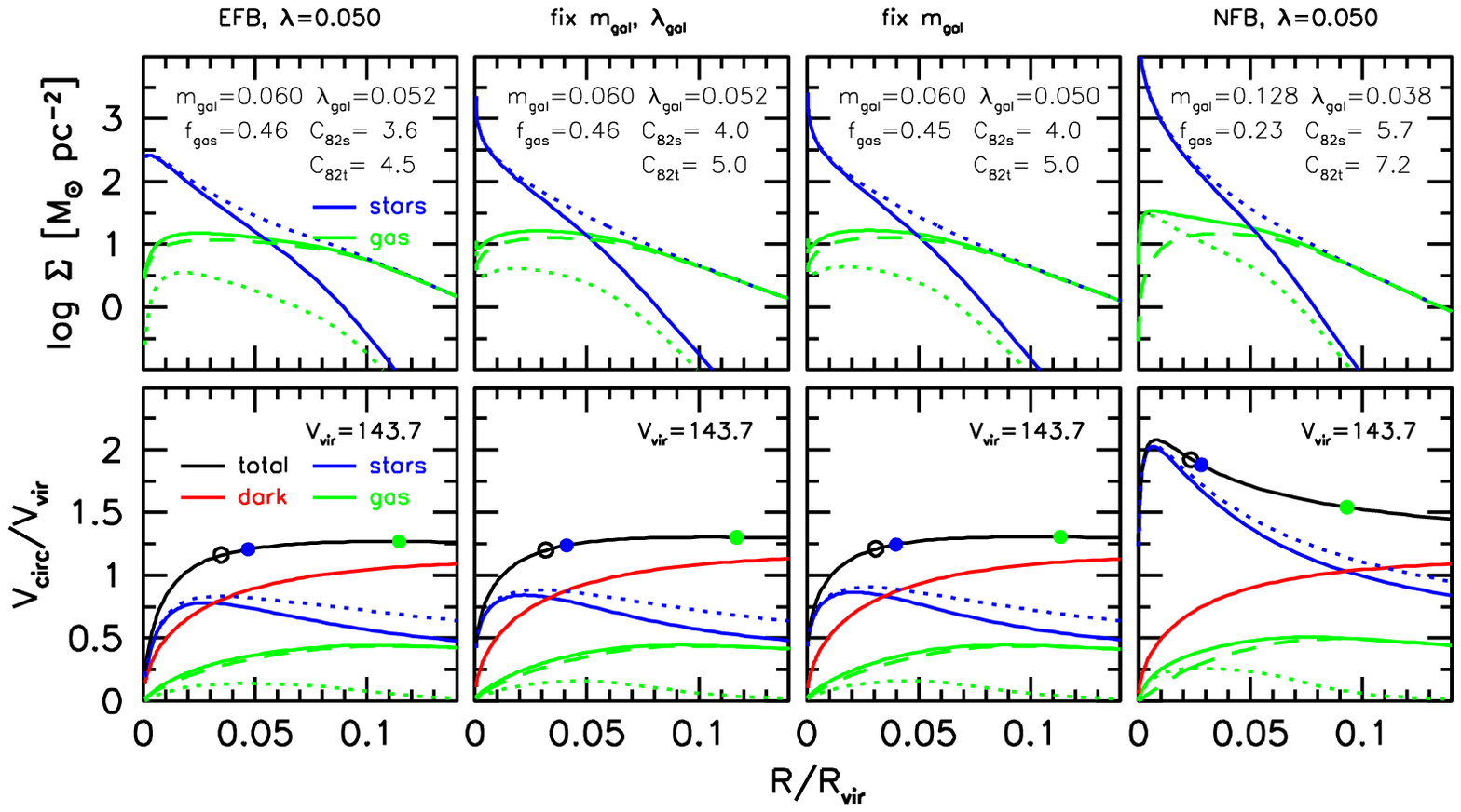,width=0.67\textwidth}}
\caption[]{\footnotesize Feedback modifies the baryon fraction, total
  angular momentum and angular momentum distribution of the baryons.
  The models on the far left are the $\Mvir=10^{10}\hMsun$ (upper panels)
  and $\Mvir=10^{11}\hMsun$ (middle panels) and and
  $\Mvir=10^{12}\hMsun$ (lower panels) models from
  Fig.\ref{fig:sv-exp} (which were chosen to have exponential stellar
  disks).  These models have energy driven feedback. The models second
  from the left have been constructed to have the same $\mgal$ and
  $\lamgal$ as the models on the far left. These models have different
  baryonic and stellar disk density profiles, with higher
  concentrations and smaller stellar disks in the models without
  feedback.  This shows that feedback is modifying the angular
  momentum distribution. The models second from the right also have
  $\mgal$ fixed to the value obtained with feedback, but has $\lamgal$
  fixed to the $\lambda$ of the halo.  The models on the far right have no
  feedback, and thus the $\mgal$ and $\lamgal$ are determined by the
  efficiency of cooling. These models have too many baryons, which
  results in disks that are too small, too concentrated, and with
  unrealistic rotation curves. See text for further details.}
\label{fig:sv6}
\end{figure*}

\subsection{Are Outflows Needed to produce Exponential Stellar Disks?}
We have shown that a model with SN driven mass outflows can
reproduce the observed dependence of light concentration with stellar
mass of spiral galaxies. In particular this model can produce pure
exponential stellar disks over a wide range in halo masses
($\Mvir=10^{10}-10^{12} \hMsun$).  We now ask the question of whether
this result requires mass outflows, or whether it could be achieved by
a model in which the baryon fraction was regulated by another
mechanism. In another words: Is getting the right amount of cold
baryons in a galaxy sufficient to produce exponential stellar disks?

We test this hypothesis by running models with the same galaxy mass
fraction at redshift zero as our exponential stellar disk models from
Fig 12.  We implement this in our model by turning off the cooling
time calculation, and instead we set the amount of baryons that
accrete onto the disk, at each time step, to be the desired galaxy
mass fraction. The result is a model with our desired galaxy mass
fraction as well as a galaxy spin parameter equal to the input halo
spin parameter.  We run this procedure for three halo masses:
$\Mvir=10^{10},10^{11}$, \& $10^{12} \hMsun$. The results of these test
are shown in Fig.~\ref{fig:sv6}. 

The models with feedback (and exponential stellar disks) are on the
far left. Models with the same galaxy mass fractions and galaxy spin
parameters (set by hand) are in the panels second from the left.
These models have smaller disks with higher concentrations. This
demonstrates that outflows have modified the distribution of angular
momentum of the baryons relative to that of the dark matter. The
models second from the right have galaxy mass fractions fixed to those
of the models with feedback, but with galaxy spin parameters set equal
to that of the dark matter.  The models on the right have no feedback,
and thus the $\mgal$ and $\lamgal$ are determined by the efficiency of
cooling, which is close to 100\% for these halo masses. These models
have too many baryons, which results in disks that are too small, too
concentrated, and with unrealistic rotation curves.

This sequence of models shows that getting the right amount of baryons
into a galaxy is not sufficient to produce exponential stellar
disks. Furthermore, getting the right amount of baryons and specific
angular momentum is also not sufficient to produce exponential stellar
disks. For the models to produce exponential stellar disks they need
to get the right amount of baryons, specific angular momentum, and
distribution of specific angular momentum.  SN driven mass
outflows are a mechanism that reduces galaxy mass fractions and at
the same time naturally modifies the angular momentum distribution of
the baryons in the desired way.  It remains to be seen whether other
mechanisms for reducing the baryon fraction of galaxies, can be found
that also modify the angular momentum of baryons in the desired way.

\section{Discussion}
\subsection{Are Galaxy Disks Universally Exponential?}
A common assumption in galaxy formation models, and observational
bulge-to-disk decompositions is that disks are exponential.  This is
not without some motivation.  1) Galaxies that are observed to be
bulge-less (i.e. they have no spheroid), invariably have close to
exponential light profiles, NGC 300 is a example of a galaxy with an
exponential stellar disk over 10 scale lengths (Bland-Hawthorn \etal
2005).  2) Even if a galaxy is not bulge-less, the disk is often well
fitted by an exponential over several scale lengths, so it is tempting
(especially when plotting surface intensity on a log-linear plot) to
extrapolate the exponential disk to small radii, and to associate all
of the excess light to a bulge.  3) Exponential profiles are also easy
to work with analytically: mass profiles, and rotation curves can be
computed easily.  However, there are several reasons to question the
assumption that stellar disks are universally exponential.

From a  theoretical point of  view, the formation of  pure exponential
disks  is a  mystery. It  has been  known since  Mestel (1963)  that a
uniform sphere in solid body  rotation has a specific angular momentum
distribution close  to that of an  exponential disk.  But  as shown by
e.g.   Dalcanton Spergel \&  Summers (1997),  when a  gas disk  with a
solid body AMD  is in centrifugal equilibrium in  an dark matter halo,
the  center of  the disk  is significantly  more concentrated  than an
exponential, and  there is  an outer  cut off at  around 4  disk scale
lengths.

Viscous  evolution  is  often  cited  as a  mechanism  for  explaining
exponential disks  (Olivier, Blumenthal, \& Primack  1991; Ferguson \&
Clarke 2001; Bell 2002).  However, it requires initial conditions that
are less concentrated  than exponential to start from,  which in \LCDM
is {\it the}  problem.  Viscous process always increase  the amount of
mass  at small radii,  so even  if it  is a  mechanism to  explain why
stellar disks are exponential over a large number of scale lengths, it
does not explain the existence of bulge-less exponential disks.

Angular momentum distributions of baryons and dark matter in a \LCDM
cosmology are typically quite different from that required to produce
exponential disks (Bullock \etal 2001b; Sharma \& Steinmetz 2005). In
particular, they tend to have too much low and high angular momentum
material. The excess of high angular momentum material may not be a
serious problem as at larger radii gas dominates over stars.  In
addition gas at large radii can be ionized making it hard to
detect. The excess of low angular momentum material is considered the
more serious problem.

We have shown that it is possible (through SN driven outflows and
density dependent star formation) to make disk galaxies with realistic
stellar surface brightness profiles with \LCDM initial conditions.  In
many cases the disk is exponential over several scale lengths, but
upturns above an exponential profile at small radii are ubiquitous.
When present in observed galaxies, these upturns above an exponential,
are generally considered to be due to a bulge or a pseudo-bulge.
However, given that there is no theoretical motivation for the
existence of pure exponential disks, it seems reasonable to suppose
that in some cases these upturns are actually part of the disk. Such a
conclusion has been reached before: quoting from Kormendy \& Kennicutt
2004 {\it ``We already accept Freeman 1970 Type II profiles as
  canonical disk behaviour, even though we can explain it in only a
  few cases (e.g.  Talbot, Jensen \& Dufour 1979).  We accept outer
  cutoffs (e.g.  van den Kruit \& Searle 1981a,b, 1982).  Oval disks
  are only piece-wise exponential.  Would it be a surprise if disks
  also knew how to deviate above an exponential to small radii?''}

\subsection{On the Origin of the Hubble Sequence}

Observationally it is well established that galaxy concentration (e.g.
Kauffmann \etal  2003); bulge fraction  (e.g.  Benson \etal  2007) and
S\'ersic  index (e.g.   this paper)  increase with  stellar  mass.  An
alternative  way  to  frame  these  observations is  that  the  Hubble
sequence   of   galaxy   types   is   also,   on   average,   a   mass
sequence. Understanding  the origin of  the Hubble sequence is  one of
the most  fundamental problems in astrophysics.  Here  we identify four
mechanisms   that  give  plausible   explanations  for   the  observed
trends. All  of these  mechanisms likely occur,  but it remains  to be
determined  which, if any,  process dominates  on any  particular mass
scale.

1) {\bf Major  Merger Rate}: Based on the differences  in the shape of
the halo  mass function to that  of the stellar  mass function, Maller
(2008) showed that major galaxy mergers are more common in higher mass
galaxies.  For  example, the largest  contribution in mass  comes from
mergers with  a mass ratio of  1:10 (e.g.  Stewart \etal  2008).  In a
halo with mass less than  about $10^{12}\Msun$ a 1:10 halo mass merger
ratio will result in a less than 1:10 galaxy mass merger ratio, whereas
in a  halo of mass greater  than about $10^{12}\Msun$ a  1:10 halo mass
merger  ratio will result  in a  greater than  1:10 galaxy  mass ratio.
Since  major  mergers   destroy  stellar  disks  producing  spheroidal
remnants (e.g. Barnes 1992; Cox  \etal 2006), higher mass galaxies are
expected to have higher bulge fractions.

2)  {\bf Gas  Fractions}: It  is observationally  known that  cold gas
fractions are lower in higher  mass galaxies (McGaugh \& de Blok 1997;
Kannappan  2004). For  disk  galaxies  this is  likely  caused by  the
combined effects of the Schmidt  law for star formation (in which star
formation is less efficient at lower gas densities) and the effects of
outflows (which result in lower density disks in lower mass galaxies).
Theoretically  it is  known  that  the bulge  fraction  of the  merger
remnant  increases with  decreasing  gas fraction  of the  progenitors
(e.g. Springel \& Hernquist  2005; Robertson \etal 2006; Hopkins \etal
2009).   Thus we  expect higher  mass  galaxies to  have higher  bulge
fractions due to major and intermediate mass ratio mergers.

3)  {\bf  Secular  Evolution}:  Secular  (i.e. slow  relative  to  the
dynamical time of the system) processes such as bars, oval distortions
and  spiral  structure are  known  to  redistribute  mass within  disk
galaxies  (see Kormendy  \& Kennicutt  2004  for a  review). A  simple
criterion  for   the  onset  of  disk  instability   is  the  relative
contribution of  the disk  to the total  rotation velocity  within the
optical radius  of the galaxy  (e.g.  Efstathiou, Lake,  \& Negroponte
1982).   Since the  contribution of  the  disk to  the total  rotation
velocity  increases with  stellar  mass and  surface brightness  (e.g.
Zavala  \etal 2003;  Pizagno \etal  2005; McGaugh  2005;  Dutton \etal
2007), more massive  disks are, on average, expected  to be more prone
to disk  instabilities. Thus more massive galaxies  should have larger
bulge fractions and light concentrations than less massive galaxies.

4)  {\bf  Feedback}:  In  our  model  the structure  of  the  disk  is
determined  by conservation  of specific  angular momentum.   The mass
dependence to  the disk  structure arises due  to two effects:  1) the
specific angular  momentum of galaxies increases  with decreasing halo
mass, due  to increased  outflow efficiency in  lower mass  haloes. 2)
Feedback preferentially  removes low angular  momentum material, which
directly results  in less concentrated disks.  This  model produces an
increase in the  concentration of the stellar disk  with stellar mass,
in  agreement  with observations,  but  without  secular evolution  or
mergers.
It is also worth pointing out that in order to produce galaxies with
the observed gas fractions and standard star formation laws, feedback
is required (Dutton \& van den Bosch 2008).

\subsection{Future Directions}

There are a number of avenues for future research suggested by this paper.

\subsubsection{Observations of Disk Galaxy Structure}
The observational analysis of galaxy structure that we present in \S 4
could  be  greatly  improved  upon.   The  first  step  would  be  two
dimensional S\'ersic  fits, which take  account of the  inclination of
the disk. However,  single S\'ersic fits do not  always provide a good
match to the  data, for example if there  are multiple components with
different  axis   ratios,  or  if  there  is   a  highly  concentrated
bulge/point source in  the center of the galaxy  (which will drive the
S\'ersic   index  artificially   high).   An   improvement   would  be
multi-component  fits, such  as  bulge plus  disk  (e.g. Simard  \etal
2002), or bulge plus disk plus bar fits (e.g. Weinzirl \etal 2008).

Rather than  focusing on bulge  to disk ratios, which  are notoriously
hard to determine (and depend  implicitly on how one defines the bulge
and disk), an  alternative, and perhaps more useful  approach would be
to  measure  the average  (and  dispersion  of)  de projected  surface
brightness profiles  as a  function of fundamental  galaxy parameters,
such as  stellar mass, color,  and size. Not  only would such  a study
highlight  the role  of fundamental  galaxy parameters  in determining
galaxy  structure,   but  would  also   provide  strong  observational
constraints for galaxy formation models to be tested against.

Since   disk  galaxies   often  have   strong  color   gradients,  the
concentration  of the stellar  mass may  be significantly  larger than
that  in  optical  light.   Thus  it  is  desirable  to  measure  disk
structural  parameters  in  wavelengths  that are  more  sensitive  to
stellar  mass, and  less  sensitive to  extinction  and young  stellar
populations (although these  are useful to study in  their own right).
This means near IR rest  frame wavelengths.  

The calculation  of global stellar masses  is now routine  both at low
and high  redshift.  The next step  in galaxy structure  studies is to
apply these  methods either to radial surface  brightness profiles, or
directly  to the  2-D surface  brightness maps.   The  feasibility and
importance of these procedures have been demonstrated in spiral galaxy
mass model studies  (e.g. Kranz \etal 2003; Kassin  \etal 2006).  What
is needed now is a study  of the stellar mass structural properties of
a statistical sample of galaxies.

An often ignored component of  galaxies is the neutral and atomic gas.
While the cold  gas may only contribute $\simeq  10\%$ of the baryonic
mass of Milky way mass galaxies, due to the atomic gas being typically
more extended than the stars  it is likely to contribute significantly
to the angular  momentum.  Furthermore, the gas to  stellar mass ratio
increases in  lower mass  galaxies, and in  dwarf galaxies the  gas is
often   the  dominant   component.    Thus  to   achieve  a   complete
observational picture of galaxy structure, it is essential, especially
in lower mass galaxies, to include the cold gas.

\subsubsection{Improvements to  Treatment of Angular  Momentum in Semi
  Analytic Models}
A necessary simplification  made in our model was  the assumption that
the spin  and AMD  of the halo  was, for  a given halo,  constant with
redshift.  The  validity of  this assumption needs  to be  tested with
cosmological simulations. 

We  have  shown  that a  subset  of  the  AMDs found  in  cosmological
simulations can  produce close to exponential disks,  once the effects
of star formation  and feedback are taken into  account.  An important
question  is  whether  the   merger  histories  of  these  haloes  are
consistent  with the  survival  of  a bulge-less  disk.   In order  to
address this  question one needs  to extend the disk  evolution models
discussed in this paper to  full halo merger histories (e.g.  Stringer
\& Benson 2007).   However, in order to treat  the build-up of angular
momentum self consistently we need to  know how the AMD of the halo is
correlated to its  merger history. For example we  would like to split
the AMD into  that due to major mergers  (which will generally destroy
disks), and that due to minor mergers and smooth accretion (which will
generally result in growth to the disk).

\subsubsection{Initial Conditions for Secular Evolution/Merger Simulations}
Numerical simulations of mergers and secular evolution are generally
(but not always e.g. Debattista \etal 2004) set up assuming the disk
is exponential (e.g.  Springel \& Hernquist 2005; Cox \etal 2006;
Debattista \etal 2006; Foyle, Courteau, \& Thacker 2008). The surface
density profiles of our models, which are motivated by cosmological
initial conditions, are in general not exponential. If these are used
as initial conditions, do they result in any testable differences in
galaxy structure, compared to exponential disk initial conditions.
Furthermore, do secular processes, or mergers preserve or modify the
stellar concentration-mass relation produced by our simple disk
evolution models?


\section{Summary}
\label{sec:conclusion}

We have used  a disk galaxy evolution model  to investigate the origin
of the  surface density profiles  of disk galaxies. Our  model follows
the accretion,  cooling and ejection  of baryonic mass  inside growing
dark matter haloes.  Contrary to most models of disk galaxy formation,
we do  not assume that the disk  is an exponential. In  our models the
surface  density profile  of the  disk is  determined by  the specific
angular  momentum  distribution  (AMD)   of  the  cooled  baryons  and
centrifugal equilibrium.  For the specific angular momentum of the hot
gas  we  use fitting  functions  from  cosmological simulations.   The
galaxy mass  fractions $\mgal$ are  determined by the  efficiencies of
cooling  and  feedback,  and  the  stellar  mass-to-light  ratios  are
determined by the star formation history and chemical evolution of our
models.  We have  used observations of S\'ersic indices  from the SDSS
to constrain our feedback  models and provide observational constraints
on the frequency of bulge-less  galaxies.  We summarize our results as
follows:

\begin{itemize} 

\item We  have compared  two fitting formulae  for the halo  AMD: from
  Bullock \etal  (2001b), and Sharma \& Steinmetz  (2005).  In general
  terms  these  two  formulae   give  similar  results  for  the  disk
  structure, but there are  significant differences, especially at the
  tails  of the  distributions.  It  would be  desirable  to determine
  which if either of these  AMD's provides a better description of the
  AMD's  in  cosmological  simulations,  and  whether  an  alternative
  fitting function provides a better overall fit.

\item Exponential and quasi-exponential  stellar disks can be produced
  by  our model  through a  combination of  SN-driven galactic
  outflows   (which  preferentially   remove   low  angular   momentum
  material), intrinsic variation  in the angular momentum distribution
  of the  halo gas,  and the inefficiency  of star formation  at large
  radii.

\item  We use observations  from the  SDSS NYU-VAGC  to show  that the
  median  S\'ersic index of  both red  and blue  galaxies is  a strong
  function  of stellar  mass.  For  blue galaxies,  low  mass ($\Mstar
  \simeq 10^{9}\Msun$)  galaxies have  $n\simeq 1.3$, while  high mass
  ($\Mstar\simeq  10^{11}\Msun$)  galaxies have  $n\simeq  4$, with  a
  transition  mass  of $\Mstar  \simeq  2.5\times 10^{10}\Msun$. 

\item Our model with energy driven outflows correctly reproduces the
  observed relation between S\'ersic index and stellar mass, whereas
  our models with momentum driven outflows and no outflows over
  predict the S\'ersic indices in low mass galaxies. Thus the
  exponential density profile is not a trivial result of cosmological
  initial conditions.

\item The observed fraction  of ``bulge-less'' exponential galaxies is
  a  strong function  of stellar  mass.  For  Milky-Way  mass galaxies
  ($\Vrot  \simeq 220  \kms,  \Mstar \simeq  10^{11}\Msun$) less  than
  0.1\% of blue galaxies are bulge-less, whereas for M33 mass galaxies
  ($ \Vrot \simeq 120 \kms, \Mstar \simeq 10^{10}\Msun$ bulge-less and
  quasi-bulgess galaxies  are more common  with $\simeq 45\%$  of blue
  galaxies having S\'ersic index  $n<1.5$.  These results suggest that
  the   difficulty  of  hierarchical   formation  models   to  produce
  bulge-less  Milky-Way   mass  galaxies  (e.g.    Abadi  \etal  2003;
  Robertson \etal  2004; Okamoto \etal 2005; Governato  \etal 2007) is
  in  fact  not  a  problem  for $\LCDM$.   However,  the  problem  of
  producing galaxies like M33 remains, and will provide a key test for
  hierarchical galaxy formation models,  and in particular the role of
  feedback.

\item Contrary to general assumptions, our models suggest that galaxy
  disks do not have a universal exponential density profile,
  especially at small radii.
\end{itemize}

\section*{Acknowledgements} 

I thank, Frank van den Bosch, Eric Bell, St\'ephane Courteau, Avishai
Dekel, Sandy Faber, Neil Katz, Andrea Macci\`o, and Julio Navarro for
useful discussions. I thank the referee for providing useful
suggestions for improving the manuscript.  A.A.D.  acknowledges
support from the National Science Foundation Grants AST-0507483,
AST-0808133.

Funding for the  Sloan Digital Sky Survey (SDSS)  has been provided by
the Alfred  P. Sloan  Foundation, the Participating  Institutions, the
National  Aeronautics and Space  Administration, the  National Science
Foundation,   the   U.S.    Department   of   Energy,   the   Japanese
Monbukagakusho,  and the  Max Planck  Society.  The SDSS  Web site  is
http://www.sdss.org/.

The SDSS is managed by the Astrophysical Research Consortium (ARC) for
the Participating Institutions. The Participating Institutions are The
University of Chicago, Fermilab, the Institute for Advanced Study, the
Japan Participation  Group, The  Johns Hopkins University,  Los Alamos
National  Laboratory, the  Max-Planck-Institute for  Astronomy (MPIA),
the  Max-Planck-Institute  for Astrophysics  (MPA),  New Mexico  State
University, University of Pittsburgh, Princeton University, the United
States Naval Observatory, and the University of Washington.


\label{lastpage}

\end{document}